\documentclass{elsart}
\usepackage{graphicx,amsmath,amssymb,amsfonts}

\begin{document}

\begin{frontmatter}



\title{Darwin-Vlasov Simulations of magnetized Plasmas}


\author{H. Schmitz} and
\author{R. Grauer}

\address{Theoretische Physik I, Ruhr-Universit\"at Bochum, 44780 Bochum,
Germany}

\begin{abstract}
% Text of abstract
  We present a new Vlasov code for collisionless plasmas in the
  nonrelativistic regime. A Darwin approximation is used for
  suppressing electromagnetic vacuum modes. The spatial integration is
  based on an extension of the flux-conservative scheme, introduced by
  Filbet et al. \cite{FIL01}. Performance and accuracy is demonstrated
  by comparing it to a standard finite differences scheme for two test
  cases, including a Harris sheet magnetic reconnection scenario. This
  comparison suggests that the presented scheme is a promising
  alternative to finite difference schemes.
\end{abstract}

\end{frontmatter}

\section{Introduction}

For many interesting problems in plasma physics a deeper understanding can
only be gained with the help of kinetic plasma models. In contrast to
macroscopic models where only the streaming velocity, pressure,
temperature and other fluid quantities are considered, the kinetic
description deals with the particle distribution function $f(\mathbf{x},
\mathbf{v})$. Here $\mathbf{x}$ is the position in space and
$\mathbf{v}$ is the three dimensional velocity. The equation describing
the evolution of the six dimensional distribution function in phase space
for a collisionless system is the Vlasov equation. The complexity of
kinetic plasma descriptions arises because of the electromagnetic fields
appearing in the Vlasov equation. These depend on the charge density and
current density in the plasma which are in turn given by the moments of
the distribution function. Thus the system presents itself as a strongly
coupled non--linear system of partial differential equations. Analytical
solutions of this system are only possible for a very restricted number of
special problems.

Among the numerous problems, where collisionless kinetic plasma
simulations are important, we name only two topics, which are important in
both laboratory- and astrophysical settings. These are the thin current
sheet region in collisionless magnetic reconnection processes and the
structure of collisionless shocks. The Harris sheet model \cite{HARS62}
gives a kinetic equilibrium of a current sheet separating two regions of
different magnetic field. While this can be formulated analytically, the
reconnection process, which is initiated by small perturbations from the
Harris equilibrium, is not yet fully understood. Also, there has been
tremendous progress in the last 10 years, both on the fluid level (see
\cite{Biskamp-etal:1997,Lottermoser-Scholer:1997,Shay-Drake:1998,Shay-etal:2001,Wang-etal:2001,Huba-Rudakov:2004}) 
and using kinetic simulations (see
\cite{Buechner-Kuska:1999,Horiuchi-Sato:1999,Pritchett:2001,Pritchett:2001b,Hesse-etal:2001,Sydora:2001,Ricci-etal:2002a,Ricci-etal:2002b,Zeiler-etal:2002,Scholer-etal:2003}
for Particle in Cell simulations and  \cite{SIL03,WIEG01} for Vlasov
simulations). Still many questions remain to be solved, such as: the
spontaneous onset of reconnection, three dimensional effects, turbulence
in the reconnection zone, acceleration of particles, and comparison to
experiments. It has become clear that, especially for the acceleration of
particles, fully kinetic models and simulations are inevitable. Up to now
most kinetic simulations make use of the Particle in Cell method (PIC), and
very few deal with a direct integration of the Vlasov equation. In order
to validate PIC simulations of collisionless magnetic reconnection, it is
necessary to simulate the same problems using different schemes, and Vlasov
schemes are an important alternative method.

The second important system currently investigated is the problem of
collisionless shocks. In the framework of MHD--equations these shocks
appear as singular points, where the macroscopic quantities show a
discontinuity. To understand the inner structure of shocks, kinetic
models are necessary since only they can describe the underlying
dissipation processes. For some special geometries together with low
Mach numbers there are time independent analytical solutions to the
kinetic equations, corresponding to stable shock conditions. For higher
Mach numbers these solutions break down, and a time dependant behaviour
appears. This leads to excitation of wave modes with a possibility of
particle acceleration. This time dependant behaviour is not accessible
with analytical methods, and numerical approaches have to be attempted.

As mentioned above, in practise there are two different numerical
approaches to solving the Vlasov equation. In the Particle in Cell
(PIC) approach trajectories of individual representative particles are
followed through the electromagnetic fields. The fields are given on a
numerical grid, while the positions and velocities of the simulation
particles can be any value.  Due to the finite number of particles,
the PIC method suffers from considerable numerical noise. A related
problem is the fact that only those parts of the distribution function
can be calculated with high precision which contain many particles in
a phase space volume. In particular the high energy tails of the
distribution function cannot be resolved.

The other numerical approach integrates the Vlasov equation directly
on a high dimensional numerical grid in phase space. These 
Vlasov schemes do not suffer from any numerical noise. The tails
of the distribution function can be modelled with high accuracy, and
deviations from a Maxwellian distribution can be pinpointed. These
advantages are traded against higher computational effort of the
Vlasov--codes as compared to PIC--codes. Additionally, care has to be
taken for choosing the integration scheme of Vlasov's equation, taking
into account the hyperbolic nature and corresponding integral
quantities.

The distribution function has to fulfil a number of restrictions
which are derived from its physical interpretation as probability
density in phase space and from the properties of the Vlasov equation. The
probability density interpretation implies that the distribution
function has to remain positive at all times. One property of Vlasov's
equation is that the values of the distribution function are
transported along the characteristics through phase space without
modification. When starting from a positive distribution function this
not only implies that the positivity is preserved but also that both
upper and lower bounds of the distribution function remain unchanged,
and that no new maxima or minima are generated. A second property of
Vlasov's equation is the conservation of phase space density as a
consequence of Liouville's equation.

A numerical scheme cannot satisfy all the above criteria exactly.
Therefore a number of numerical schemes have been proposed, each of
which is a compromise between different requirements. Spectral codes
which solve the Vlasov equation in the Fourier domain suffer only from
little numerical diffusion, but they are mostly limited to periodic
boundary conditions, see e.g.\ \cite{ARM76}. Another important
drawback of these methods is that they do not preserve positivity,
let alone the number of extrema in the distribution function.
Eulerian solvers, on the other hand, allow for non--periodic boundary
conditions and can be made to preserve positivity and the maximum
principle. They are, however, slightly more diffusive than spectral
codes. A recent comparison of Eulerian solvers can be found in
\cite{ARB02}. In this work we will use a flux conservative and
positive scheme \cite{FIL01} which obeys the maximum principle and
suffers from relatively little numerical diffusion.

The integration of the Vlasov equation has to be performed
simultaneously to the evaluation of the Maxwell equations. The
dynamics of the full electrodynamic fields imposes an additional
criterion for the time step used in the simulation. Since
electromagnetic waves can travel through the system, the time step has
to be chosen such that the speed of light $c$ is resolved on the
numerical grid: $\Delta t < c \Delta x$, where $\Delta t$ is the time
step and $\Delta x$ is the grid resolution. One way to avoid this is
to use an electrostatic model. This is, however, only applicable in
special situations where the self generated magnetic field can be
neglected. Here we will use the Darwin approximation of Maxwell's
equations, which follows from a rigorous expansion of the full Maxwell
equations in orders of $v^2 / c^2$, where $v$ is a characteristic
velocity of the system. In the framework of the Darwin approximation,
the purely electromagnetic modes are neglected but electrostatic,
magnetostatic, and inductive fields are still considered. Darwin's
approximation has been widely used with Particle in Cell simulations
\cite{NIE76,BIR85}, however, it has not found its way into Vlasov
simulations yet. An alternative method is the use of an implicit time
stepping as it has been implemented by \cite{Ricci-etal:2002a} in
the Celested3D code.

The next section will present the basic equations together with their
normalisation. In section \ref{SecDarwin} we will introduce the Darwin
approximation of the Maxwell equations. Section \ref{SecFluxCons} will
give an overview of the one dimensional flux conservative scheme used for
integrating Vlasov's equation, while section \ref{SecCharacter} will
describe the time splitting schemes that provides the generalisation to the
5--dimensional phase space. In section \ref{SecResults} results are
presented, and section \ref{SecConclusion} contains some concluding remarks.

\section{Basic equations}

In this and the next section we want to present the basic equations and
approximations of our model. The aim is to simulate Vlasov's equation
\begin{displaymath}
\frac{\partial f_k}{\partial t} 
        + {\bf v}\cdot \nabla f_k
        + \frac{q_k}{m_k} \left( 
                {\bf E} + {\bf v} \times {\bf B}
        \right) \cdot \nabla_{\bf v} f_k 
= 0.\label{VlasovOrig}
\end{displaymath}

Here $f_k({\bf x}, {\bf v},t)$ is the distribution function of species
$k$. In this work only singly charged ions and electrons $k = i,e$ are
considered, although the code allows arbitrary species. The quantities
$q_k$ and $m_k$ are the charge and the mass of the particles of species
$k$. The Lorentz force depends on the electric and magnetic fields. These
are in general given by Maxwell's equations,
\begin{alignat}{1}
  \nabla\times\mathbf{E} &= -\frac{\partial \mathbf{B}}{\partial t} \;\; ,\\
  \frac{1}{\mu_0}\nabla\times\mathbf{B} &=
  \varepsilon_0 \frac{\partial \mathbf{E}}{\partial t} + {\bf j} \;\; ,\\
  \nabla\cdot\mathbf{E} &= \frac{1}{\varepsilon_0} \rho \;\; ,\\
  \nabla\cdot\mathbf{B} &= 0 \;\; ,
\end{alignat}
where $\mathbf{B}$ is the magnetic and $\mathbf{E}$ the electric
field. In section \ref{SecDarwin} we will present the Darwin
approximation of Maxwell's equations, which is used to solve the
electromagnetic fields in our simulation code. The charge density
$\rho$ and the current density $\mathbf{j}$ are given by the moments
of the distribution function,
\begin{alignat}{1}
\rho &= \sum_k q_k \int f_k(\mathbf{x}, \mathbf{v}) d^3v \;\; ,\\
\mathbf{j} &= \sum_k q_k \int \mathbf{v} f_k(\mathbf{x}, \mathbf{v}) d^3v \;\; .
\end{alignat}

\subsection{Normalisation}

Here we want to present the normalisation of the Vlasov--Maxwell system
equations. For this we introduce normalising parameters $A_0$, where $A$
stands for any of the physical quantities. The normalised quantities
$\hat{A}$ are then given by $\hat{A} = A / A_0$.

The un--normalised characteristics of Vlasov's equation for species $k$
are given by
\begin{alignat}{1}
\frac{d\mathbf{x}}{dt} &= \mathbf{v} \;\; ,\\
\frac{d\mathbf{v}}{dt} &= \frac{q_k}{m_k}\left( \mathbf{E}+\mathbf{v}\times
\mathbf{B} \right) \;\; .
\end{alignat}
Here we have used the symbols $\mathbf{x}$ and $\mathbf{v}$ to denote the
characteristics $\mathbf{x}=\mathbf{x}(t,\mathbf{x}_0, \mathbf{v}_0)$ and
$\mathbf{v}=\mathbf{v}(t,\mathbf{x}_0, \mathbf{v}_0)$. These have to be
distinguished from the independent variables $\mathbf{x}$ and $\mathbf{v}$
in Vlasov's equation. In the following the appropriate meaning should be
clear from the context.
The form of the above equations is not modified by the normalisation,
\begin{alignat}{1}
\frac{d\hat{\mathbf{x}}}{d\hat{t}} &= \hat{\mathbf{v}} \;\; ,\\
\frac{d\hat{\mathbf{v}}}{d\hat{t}} &= \frac{\hat{q}_k}{\hat{m}_k}
\left( \hat{\mathbf{E}} + \hat{\mathbf{v}}\times\hat{\mathbf{B}} \right) \;\; .
\end{alignat}

Only the individual charge to mass ratios $\hat{q}_k/\hat{m}_k$ are
relevant parameters of the system. On the other hand, Maxwell's
equations simplify to
\begin{alignat}{1}
\nabla\times \hat{\mathbf{E}} &= - \frac{\partial \hat{\mathbf{B}}}{\partial \hat{t}} \;\; ,\\
\nabla\times \hat{\mathbf{B}} &=
    \alpha^2\left( \frac{\partial \hat{\mathbf{E}}}{\partial \hat{t}} +
                \hat{\mathbf{j}} \right) \;\; ,\\
\nabla \cdot \hat{\mathbf{E}} &= \hat{\rho} \;\; ,\\
\nabla \cdot \hat{\mathbf{B}} &= 0 \;\; .
\end{alignat}
Here $\alpha=v_0 / c$ is the ratio of the normalisation velocity over the
speed of light.

We choose $m_0=m_i$ and $q_0=e$, where $m_i$ is the ion mass and $e$ is
the unit electron charge. $B_0$ and $n_0$ remain free to choose. Then
$x_0 = \lambda_i$ is the ion inertial length, $t_0 = 1/ \Omega_{i}$ is
the inverse ion gyro frequency ($\Omega_i = \frac{e B}{m_i}$), and $v_0
= v_A = B_0 / \sqrt{\mu_0 m_i n_0}$ is the Alfv\'en velocity.

We still have the freedom of choosing the magnetic field $B_0$. If we
choose $B_0$ to be a characteristic magnetic field magnitude in the
system, then $\alpha$ gives the ratio of Alfv\'en velocity to the speed of
light in the system. On the other hand we can choose $B_0$ such that
$\alpha=1$. Then the local magnetic field magnitude in the system
determines $v_A / c$. In this work we will choose the first, so that the
field magnitude in the system remains around unity and the free parameter
$\alpha$ can be used to set the magnetic field strength.  In the following
we will drop the hat--notation ($\hat{\;}$) for the normalised quantities.

\section{Darwin approximation\label{SecDarwin}}

The flux conservative integration scheme of the Vlasov equation, which will
be presented in section \ref{SecFluxCons}, is not restricted by a CFL
condition on the time step. However, when combined with the Maxwell
equations for the electromagnetic fields a CFL condition is introduced by
the time integration of the fields on the numerical grid. This implies
that the fastest electromagnetic wave mode, i.e. the vacuum mode, has to be
resolved on the grid,
\begin{equation}
\Delta t < \frac{\Delta x}{c} \;\; ,
\end{equation}
where $c$ is the speed of light. This condition imposes severe
restrictions on the time step. In many applications the
electromagnetic vacuum modes are not important. The standard solution
to this problem is the electrostatic approximation. In this
approximation only Poisson's equation needs to be solved for the
electric field. A magnetic vacuum field can also be superimposed. The
influence of the plasma on the magnetic field is, however, completely
neglected. To account for this influence, and thus the possibility of
magnetosonic wave modes, the Darwin approximation is commonly used in
particle simulations. This approximation can be derived from the
Maxwell equations in an expansion in orders of $v^2 / c^2$, where $v$
is some characteristic velocity. Assuming $v^2\ll c^2$ this leads to a
set of equations where the vacuum modes are eliminated but all other
wave modes are retained. Darwin's approximation can be used when the
velocities are small compared to the speed of light and there is no
energy transported by the electromagnetic radiation. Some applications
are the study of magnetic reconnection, e.g. in the earths magnetotail
\cite{Pritchett:1991}, or the investigation of high intensity charged
particle beams \cite{Lee:2001}.  Naturally it is not capable of
describing phenomena where the electromagnetic vacuum modes play a
major role, e.g. in laser--plasma interaction.

Darwin's approximation starts from a separation of the electric field into
a longitudinal and a transverse part \cite{BIR85}
\begin{equation}
\mathbf{E} = \mathbf{E}_L + \mathbf{E}_T \;\; ,
\end{equation}
with
\begin{equation}
\nabla \times \mathbf{E}_L = 0 
\quad \text{and} \quad
\nabla \cdot \mathbf{E}_T = 0 \;\; . \label{ESplit}
\end{equation}
 
The normalised Maxwell's equations are then approximated by
\begin{alignat}{1}
\nabla\cdot\mathbf{E}_L  &= \rho \;\; ,\\
\nabla\cdot\mathbf{B}    &= 0 \label{DivBZero}\\ %\phantom{\frac{1}{1}}\;\; ,\\
\nabla\times\mathbf{E}_T &= - \partial_t \mathbf{B} \label{erregE}\;\; ,\\
\nabla\times\mathbf{B}   &= \alpha^2
        \left(  
                \partial_t \mathbf{E}_L + \mathbf{j}
        \right) \label{erregB} \;\; .
\end{alignat}
The approximation only appears in eq (\ref{erregB}), where only the
longitudinal part of the displacement current is taken. All other
equations remain unchanged. In vacuum, equations (\ref{erregE}) and
(\ref{erregB}) are now decoupled and no purely electromagnetic modes can
appear.

An advantage of using Darwin's approximation instead of the full Maxwell
equations is the fact, that the equations can be solved without
performing a time integration step. All the electromagnetic
fields can be calculated from the moments of the distribution at a given
time together with the boundary conditions for the fields. This will be
presented in the following. 

Poisson's equation for the electrostatic potential
\begin{displaymath}
\Delta\Phi = -\rho 
\quad \text{with} \quad
\mathbf{E}_L = -\nabla \Phi
\end{displaymath}
immediately gives the longitudinal electric field.  Taking the curl of
(\ref{erregB}), together with (\ref{ESplit}) and (\ref{DivBZero},
results in
\begin{displaymath}
\Delta \mathbf{B} = -\alpha^2\nabla\times\mathbf{j} \;\; ,
\end{displaymath}
giving three Poisson equations for the components of the magnetic field.
To calculate $\mathbf{E}_T$ the curl of eq (\ref{erregE}) is taken and eqs
(\ref{erregB}) and (\ref{ESplit}) substituted, giving
\begin{alignat}{1}
\Delta \mathbf{E}_T &= \partial_t \nabla\times\mathbf{B}\\
&= \alpha^2 \partial_t \mathbf{j} - \nabla \left( \alpha^2 \partial_{tt}\phi \right) \;\; .
\label{ETransOrig}
\end{alignat}
To eliminate the time derivative of the current density,
$\mathbf{j}$ is expressed as moment of the distribution function,
\begin{displaymath}
\partial_t \mathbf{j} = \sum_k q_k \int \mathbf{v}\partial_t f_k\; d^3v.
\end{displaymath}
Here the index $k$ sums over all particle species. Now, Vlasov's equations 
for the different species are substituted,
\begin{displaymath}
\partial_t \mathbf{j} = 
- \sum_k \nabla\rho_k \langle \mathbf{v v} \rangle_k
+ \sum_k \frac{q_k \rho_k}{m_k} \mathbf{E}
+ \sum_k \frac{q_k \rho_k}{m_k} \langle \mathbf{v} \rangle_k \times
        \mathbf{B} \;\; .
\end{displaymath}
Here the pointed brackets $\langle . \rangle_k$ denote averaging with the
distribution function $f_k$.
The electric field appearing on the right hand side is, of course, comprised
of the longitudinal and the transverse part. While the longitudinal
component is already known from the charge density, it is the transverse
component that is being calculated here. Inserting $\partial_t \mathbf{j}$
into eq (\ref{ETransOrig}) and introducing the local  plasma frequency
\begin{equation}
\omega^2 = \sum_k \frac{q_k \rho_k}{m_k} \;\; ,
\end{equation}
this leads to
\begin{multline}
\Delta \mathbf{E}_T - \alpha^2\omega^2 \mathbf{E}_T = \\
        - \sum_k \nabla\rho_k \langle \mathbf{v v} \rangle_k
+ \sum_k \frac{q_k \rho_k}{m_k} \mathbf{E}_L
+ \sum_k \frac{q_k \rho_k}{m_k} \langle \mathbf{v} \rangle_k \times
        \mathbf{B}
- \nabla \left( \alpha^2 \partial_{tt}\phi \right) \;\; .
\end{multline}
This equation is a Helmholtz--equation for each component of
$\mathbf{E}_T$. The last term on the right hand side can be deduced
from the condition $\nabla \cdot \mathbf{E}_T = 0$, since it only adds
a curl free component to $\mathbf{E}_T$. It can thus be omitted to
calculate $\mathbf{\tilde{E}}_T$, with
\begin{displaymath}
\Delta \mathbf{\tilde{E}}_T - \alpha^2\omega^2 \mathbf{\tilde{E}}_T =
        - \sum_k \nabla\rho_k \langle \mathbf{v v} \rangle_k
+ \sum_k \frac{q_k \rho_k}{m_k} \mathbf{E}_L
+ \sum_k \frac{q_k \rho_k}{m_k} \langle \mathbf{v} \rangle_k \times
        \mathbf{B} \;\; .
\end{displaymath}
$\mathbf{\tilde{E}}_T$ is then projected onto its divergence free part by
calculating $\Theta$ with $\Delta \Theta = \nabla \cdot
\mathbf{\tilde{E}}_T$  to give  $\mathbf{E}_T = \mathbf{\tilde{E}}_T -
\nabla \Theta$. 

Altogether this sums up to 8 elliptic equations. However, the time spent
in solving these equations, compared to the integration of the 5 or 6
dimensional problem, is negligible. This situation is different in PIC
simulations. One has to compare about 30 particles per cell used in
standard PIC simulations to $10^3$ - $30^3$ mesh points for the resolution
of the velocity space used in Vlasov codes. For this reason the
computational effort of solving the above elliptic equations has an effect
on the total computational time in standard PIC simulations, but not in
Vlasov simulations.

\section{Flux Conservative Scheme\label{SecFluxCons}}

In this section we briefly present the numerical scheme used for
integrating the Vlasov equation. This scheme has originally been
presented in Filbet et al. \cite{FIL01}. The scheme uses a flux
conservative formulation and is based on a third order reconstruction
of the primitive of the distribution function using a fixed stencil.

To formulate the scheme we start from the observation that in Hamiltonian
systems the values of the distribution function are transported along the
characteristics
\begin{equation}
f(\xi,t) = f(X(s,t,\xi),s). \label{CharacTransport}
\end{equation}
Here $X(s,t,\xi)$ denotes the characteristic with parameter $s$ that
satisfies \mbox{$X(t,t,\xi) = \xi$}, where in general $\xi =
(\mathbf{x},\mathbf{v})$. For the rest of this section, we restrict the
calculations to one dimension. The generalisation to the higher
dimensional system will be given later. We thus assume
$\xi \in \mathbb{R}$. Then we can integrate (\ref{CharacTransport}) and
obtain a propagator from time $t^n$ to time $t^{n+1}$
\begin{equation}
\int\limits_{x_{i-1/2}}^{x_{i+1/2}} f(x,t^{n+1}) \; dx
= \int\limits_{X(t^n, t^{n+1}, x_{i-1/2})}^{X(t^n, t^{n+1},x_{i+1/2})}
f(x,t^n) \; dx. \label{CharacTransportInt}
\end{equation}

Here $x_{i-1/2}$ and $x_{i+1/2}$ are the boundaries of the numerical grid
cell $i$. The discretisation of the distribution function is now suggested
by  the above equation. The values $f_i^n$ on the numerical grid represent
the cell integrals
\begin{displaymath}
f_i^n = \frac{1}{\Delta x}\int\limits_{x_{i-1/2}}^{x_{i+1/2}} f(x,t^n) \; dx.
\end{displaymath}
We also define the flux through a cell boundary at $x_{i+1/2}$ during the
time interval $[t^n;t^{n+1}]$
\begin{displaymath}
\Phi_{i+1/2}^n = \int\limits_{X(t^n, t^{n+1}, x_{i+1/2})}^{x_{i+1/2}}
f(x,t^n) \; dx
\end{displaymath}
Then equation (\ref{CharacTransportInt}) can be expressed in the following
form
\begin{equation}
f_i^{n+1} = \Phi_{i-1/2}^n + f_i^n - \Phi_{i+1/2}^n.
\end{equation}
This equation still follows exactly from Vlasov's equation. However, the flux
$\Phi$ has to be calculated, and this requires an approximation of the
distribution function. 

In the framework of the third order positive and flux conservative scheme,
as presented by \cite{FIL01}, the primitive of the distribution function is
approximated using a four point stencil. Let
\begin{equation}
F(x,t^n) = \int_{x_0}^x f(x',t^n) \; dx',
\end{equation}
then it follows exactly that
\begin{equation}
F(x_{i+1/2},t^n) = \Delta x\sum_{k=0}^i f_k^n \equiv F_i^n .
\end{equation}
The primitive is given exactly on the cell boundaries. To approximate
$F(x,t^n)$ in the cell interval $[x_{i-1/2};x_{i+1/2}]$, the four points
$\{x_{i-3/2},x_{i-1/2},x_{i+1/2},x_{i+3/2}\}$ are used. 
Taking the derivative of the primitive, we recover the approximation of the
distribution function
\begin{equation}
\begin{split}
f_{\approx}(x) = f_i &+ \frac{\epsilon_i^+}{6\Delta x^2}
\bigl[ 
    2\left( x-x_i \right) \left( x-x_{i-3/2} \right) \\
    &+\left( x-x_{i-1/2} \right) \left( x-x_{i+1/2} \right)
\bigr] \left( f_{i+1} - f_i \right) \\
&- \frac{\epsilon_i^-}{6\Delta x^2}
\bigl[ 
    2\left( x-x_i \right) \left( x-x_{i+3/2} \right)  \\
    &+\left( x-x_{i-1/2} \right) \left( x-x_{i+1/2} \right)
\bigr] \left( f_i - f_{i-1} \right).
\end{split}
\end{equation}
Here limiters $\epsilon_i^{\pm}$ are introduced. The limiters are chosen
in such a way as to limit the values of the distribution function to a
fixed interval $0\le f_{\approx}(x) \le f_{\infty}$, where $f_{\infty}$ is
the maximum of all $f_i$. The $\epsilon_i^{\pm}$ can be written as
\begin{equation}
\epsilon_i^+ = 
\begin{cases}
\min(1;2f_i/(f_{i+1}-f_i)) & \text{if }f_{i+1}-f_i>0\\
\min(1;2(f_{\infty}-f_i)/(f_i-f_{i+1})) & \text{if }f_{i+1}-f_i<0\\
\end{cases} \label{LimitPlus}
\end{equation}
and 
\begin{equation}
\epsilon_i^- = 
\begin{cases}
\min(1;2(f_{\infty}-f_i)/(f_i-f_{i-1})) & \text{if }f_i-f_{i-1}>0\\
\min(1;2f_i/(f_{i-1}-f_i)) & \text{if }f_i-f_{i-1}<0\\
\end{cases}.\label{LimitMinus}
\end{equation}

If we determine $j$ such that
$x_{j-1/2} \le X(t^n, t^{n+1}, x_{i+1/2}) < x_{j+1/2}$, we get
\begin{equation}
\begin{split}
\Phi_{i+1/2} = \left( 1-\delta \right) \bigl[ f_j
    &+(1/6)\delta(\delta + 1)\epsilon^+ (f_{j+1}-f_j) \\
    &-(1/6)\delta(\delta - 2)\epsilon^- (f_j-f_{j-1})) \bigr] + 
    \Delta x \sum\limits_{k=\min(i,j)+1}^{\max(i,j)-1} f_k,
\end{split}
\end{equation}
where $\delta=X(t^n, t^{n+1}, x_{i+1/2}) - x_{j-1/2}$.

\section{Time splitting and integration of
characteristics\label{SecCharacter}}

The above scheme was presented in one dimension. To perform two
dimensional Vlasov--simulations including magnetic fields, the scheme has
to be generalised to a total of 5 dimensions. Califano et al. \cite{CAL01}
proposed a scheme in which the individual steps for the different
dimension are carried out using a second order time splitting scheme. The
scheme is an extension of the time splitting scheme presented by Cheng and
Knorr \cite{CHNG82}. For the two spatial directions $x$ and $y$, the
projection of the characteristics onto the axes is simply evaluated from
by the corresponding velocity components $v_x$ and $v_y$. In this way, the
two space dimensions are independent of each other and can be performed
sequentially,
\begin{equation}
T_{\mathbf{x}} = T_x T_y
\end{equation}
where $T_k$ denotes a time--integration step in one dimension with the
direction $k$. In contrast to this, the velocity components are not
independent of each other due to the magnetic field. To create a
second order scheme Califano et al. \cite{CAL01} formulated a
straightforward second order time splitting scheme,
\begin{equation}
T_{\mathbf{v}}(\Delta t) = 
        T_{vx} (\frac{\Delta t}{4}) 
        T_{vy} (\frac{\Delta t}{2}) T_{vx} (\frac{\Delta t}{4})
        T_{vz} (\Delta t) 
        T_{vx} (\frac{\Delta t}{4}) T_{vy} (\frac{\Delta t}{2})
        T_{vx} (\frac{\Delta t}{4}).
\end{equation}
This propagator in the velocity directions was then combined with the
propagator in the space directions to create the full scheme,
\begin{equation}
T_{\text{full}}(\Delta t) = 
        T_{\mathbf{x}}(\Delta t/2) 
        T_{\mathbf{v}}(\Delta t) 
        T_{\mathbf{x}}(\Delta t/2).
\end{equation}
In the following we will refer to this method as the {\em time splitting}
method.

We implemented and tested this scheme, but found it to suffer from
substantial inaccuracies, as shown in section \ref{SecGyroMotion}.
Therefore, a new scheme for the integration is proposed in this work which
we call the back substitution method. Here, the integration of Vlasov's
equation in the three velocity dimensions is still split into three
separate steps, one for each direction. The difference to the above scheme
is the way the integration of the characteristics is separated out into
the three substeps. This implies that the concrete implementation of the
scheme proposed here depends on the method of integration of the
characteristics. We will therefore quickly present Boris' scheme for
integration of characteristics in a magnetic field
\cite{BIR85,BUN67,BOR70}. Boris' scheme is widely used in Particle codes
and has the advantage that it is a second order integration scheme in
which the magnetic field does not cause a change in the kinetic energy.
The integration step is formulated as an implicit finite difference scheme
\begin{equation}
\frac{\mathbf{v}^{n+1} - \mathbf{v}^n}{\Delta t} = \frac{q}{m}
\left( 
        \mathbf{E} + \frac{\mathbf{v}^{n+1} + \mathbf{v}^n}{2} 
        \times \mathbf{B} .
\right)
\end{equation}
The electric and magnetic forces are separated,
\begin{alignat}{1}
\mathbf{v}^- &= \mathbf{v}^n + \frac{\Delta t}{2}\frac{q}{m}\mathbf{E}
\label{EqDefVPlus} ,\\
\mathbf{v}^+ &= \mathbf{v}^{n+1} - \frac{\Delta t}{2}\frac{q}{m}\mathbf{E}
\label{EqDefVMinus} ,
\end{alignat}
leading to
\begin{equation}
\frac{\mathbf{v}^+ - \mathbf{v}^-}{\Delta t} = \frac{q}{2m}
\left( \mathbf{v}^+ + \mathbf{v}^- \right) \times \mathbf{B} .
\end{equation}
The transformation from $\mathbf{v}^-$ to $\mathbf{v}^+$ is a pure
rotation with an angle $\theta$, where
\begin{equation}
\left| \tan \frac{\theta}{2}\right| = \frac{\Delta t}{2}\frac{qB}{m}.
\end{equation}
To implement this rotation the vectors $\mathbf{t}$ and $\mathbf{s}$ are
defined
\begin{equation}
\mathbf{t} = \frac{\Delta t}{2}\frac{q \mathbf{B}}{m} , \qquad 
\mathbf{s} = \frac{2\mathbf{t}}{1+t^2}. \label{EqDefTAndS}
\end{equation}
Then the rotation is performed in two steps
\begin{equation}
\mathbf{v}' = \mathbf{v}^- + \mathbf{v}^- \times \mathbf{t}
\label{EqDefVPrime}
\end{equation}
and
\begin{equation}
\mathbf{v}^+ = \mathbf{v}^- + \mathbf{v}' \times \mathbf{s}
\label{EqCalcVPlus} .
\end{equation}
This scheme supplies $\mathbf{v}^{n+1} = (v_x^{n+1}, v_y^{n+1},
v_z^{n+1})$ in terms of $\mathbf{v}^n = (v_x^n, v_y^n, v_z^n)$.

For the integration of Vlasov's equation in three dimensional velocity
space the three individual integration steps are performed in turn, thus,
the integration of the characteristics also has to be split into three
separate steps. For this, two things have to be considered. Firstly, when
integrating one component of the velocity, we take into account the shifts
of the distribution function that have already been performed. Secondly, we
then have to observe the order of integration, which is reversed with
respect to the integration of a particle trajectory. This follows from the
fact that, for the Vlasov scheme, the characteristics have to be traced
backwards in time.

For clarity, we will present the back substation method for forward
integration of the characteristics. Later, we will reverse the order of the
substeps for the backward integration needed in the Vlasov scheme. For
forward integration, first, the integration in $v_x$ is performed, as
described in the discussion of Boris' scheme above. For the integration of
the  characteristics this means $v_x^{n+1} = v_x^{n+1}(v_x^n, v_y^n,
v_z^n)$. When the integration along the $v_y$--direction is performed, in
the next step, the shift along the $v_x$--direction has already been
performed. For the integration of the characteristics this means that a
scheme calculating $v_y^{n+1} = v_y^{n+1}(v_x^{n+1}, v_y^n, v_z^n)$ is
needed. In the above notation, it is sufficient to reformulate the
integration to give $v_y^+ = v_y^+(v_x^+, v_y^-, v_z^-)$. This can be
achieved using simple algebraic manipulations and eliminating $v_x^-$. For
the last integration step along the $v_z$--direction, one can find in the
same manner a scheme giving $v_z^+ = v_z^+(v_x^+, v_y^+, v_z^-)$. The
details of this scheme are presented in appendix \ref{AppBackSubst}. The
scheme has also been implemented and it was found to be less diffusive
than the second order time splitting scheme (see section
\ref{SecGyroMotion}). As stated above, the characteristics have to be
traced backwards in time. To this end, we reverse the order of integration,
without changing the above formulas; i.e. we first shift the distribution
function in the $v_z$--direction using $v_z^+ = v_z^+(v_x^+, v_y^+,
v_z^-)$, then in the $v_y$--direction using $v_y^+ = v_y^+(v_x^+, v_y^-,
v_z^-)$, and finally in the $v_x$--direction using $v_x^{n+1} =
v_x^{n+1}(v_x^n, v_y^n, v_z^n)$.

The splitting of the integration into individual steps introduces another
problem, independent of the splitting scheme used. The limiters
$\epsilon_i^+$ and $\epsilon_i^-$ guarantee only that the distribution function
is positive and limited by $f_{\infty}$ from above for incompressible
transport equations. The Vlasov equation together with any set of
equations for the electromagnetic fields presents such a system. Due to
the separation of the integration along the different directions in
velocity space together with the forces originating from the magnetic
field, the single integration steps are, however, not incompressible. This
holds for both the splitting and the back substitution scheme. The first
direction of integration might compress the distribution function in the
$v_x$ direction, while the successive integrations decompress the
distribution function in the $v_y$ and $v_z$ directions. In the end the
total compression will always vanish. The maximum value of the
distribution function might however increase during an intermediate step.

We present two types of calculations. In the first we omit the limiter from
above completely, allowing the distribution function to rise
uncontrollably. The limiters $\epsilon_i^+$ and $\epsilon_i^-$ are then
given by
\begin{equation}
\epsilon_i^+ = 
\begin{cases}
\min(1;2f_i/(f_{i+1}-f_i)) & \text{if }f_{i+1}-f_i>0\\
1 & \text{if }f_{i+1}-f_i<0\\
\end{cases}
\end{equation}
and 
\begin{equation}
\epsilon_i^- = 
\begin{cases}
1 & \text{if }f_i-f_{i-1}>0\\
\min(1;2f_i/(f_{i-1}-f_i)) & \text{if }f_i-f_{i-1}<0\\
\end{cases}.
\end{equation}
This means, the maximum principle, that was fulfilled by the original scheme,
is now no longer satisfied. In the second type of simulation we keep the
original limiters (\ref{LimitPlus}) and (\ref{LimitMinus}) but determine
$f_{\infty}$ to be the global maximum of the distribution function after
each intermediate integration step for each direction in velocity space.

\section{Numerical tests\label{SecResults}}

The third order positive flux conservative method has been extensively
tested in one dimension and in the electrostatic limit \cite{FIL01}.
Here, we first want to present results of the integration scheme in
three dimensional velocity space with a given magnetic field. We
compare the results of the two integration schemes and a standard
second order finite difference scheme. The latter is included in the
comparison since it has been used for Vlasov simulations of
reconnection by various authors.

In the second part of this section we present results of reconnection
simulations. These have been carried out using both the flux conservative
scheme with the back substitution method and the finite
difference scheme. Results of the two are compared.

\subsection{Gyro motion\label{SecGyroMotion}}

To test the quality of the Vlasov integration scheme together with the
integration of the characteristics a simple test system was simulated. In
this test only one positively charged
species was simulated in a constant magnetic field $\mathbf{B} =
B_0\mathbf{e}_z$, with $B_0=1$. The initial
distribution function was taken to be a shifted Maxwellian
\begin{equation}
f(\mathbf{x}, \mathbf{v}) = \exp\left[ -(\mathbf{v} - \mathbf{v}_0)^2 \right] ,
\end{equation}
where $\mathbf{v}_0 = v_0 \mathbf{e_x}$ is a constant velocity in the
$x$--direction. There was no spatial variation, and so a simple gyro motion
of the thermal peak in velocity space is expected. However, due to
numerical errors this peak will change its shape. These numerical errors
only appear in the presence of a magnetic field. In the pure one
dimensional advection problem only small errors compared with the exact
solution are found.

\begin{figure}
\begin{center}
\includegraphics[width=8cm]{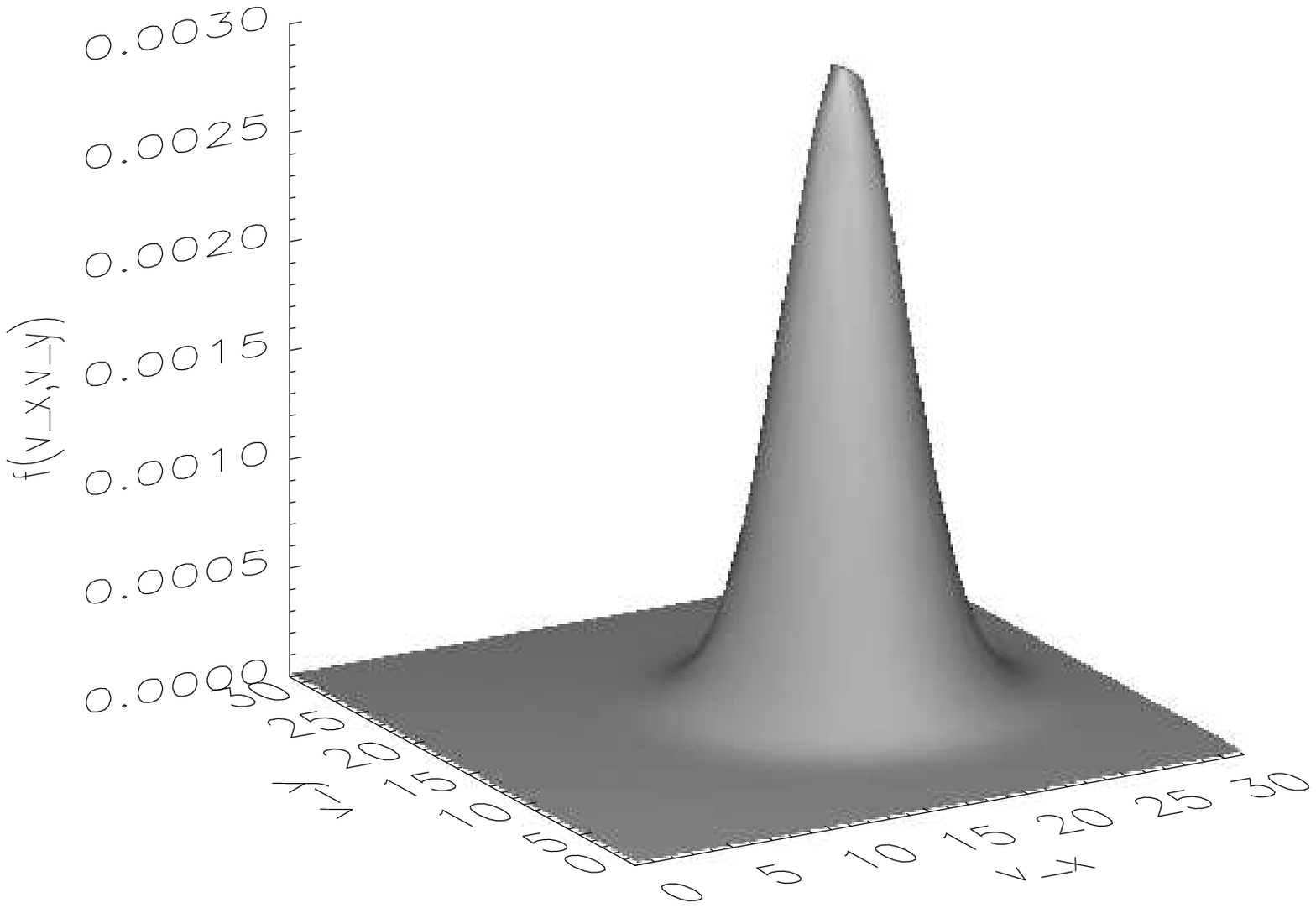}
\includegraphics[width=8cm]{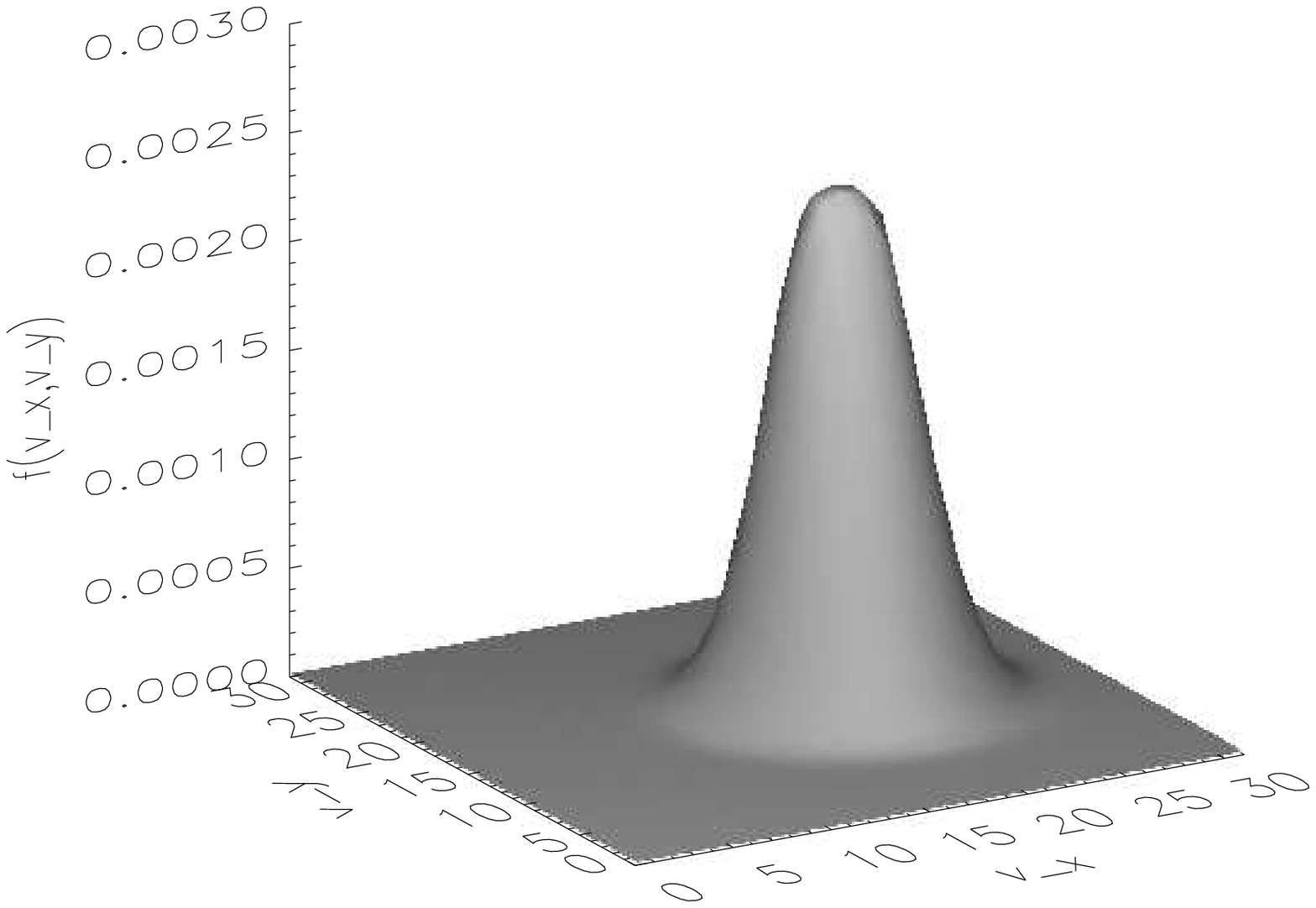}
\includegraphics[width=8cm]{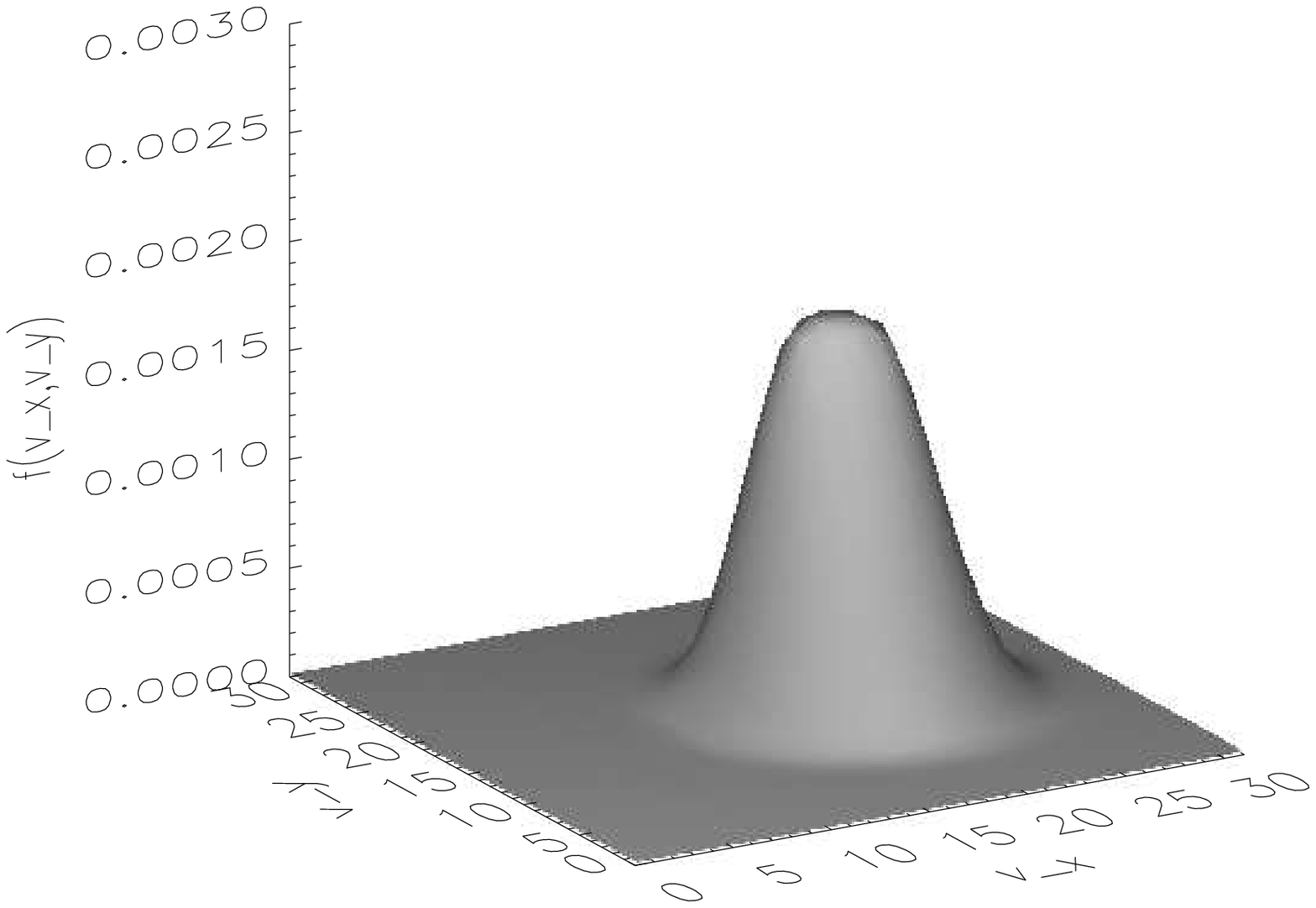}
\end{center}
\caption{Gyration of a Maxwellian peak using the back-substitution algorithm with 
$N_v=30$ and $\Delta t=2\pi\Omega_c/200$ (a) initially, (b) after one gyration 
period and (c) after five gyration periods. The scale on the $z$--axis is
in arbitrary units but identical in the three diagrams. \label{FigGyroA}}
\end{figure}

We have simulated the gyro motion of a thermal peak using the two
different integration methods for the characteristics, different time
steps, and varying resolutions of the grid in the velocity dimensions.
Additional simulations have been carried out using a simple finite
difference scheme. The scheme is obtained from eq (\ref{VlasovOrig})
by substituting all partial derivatives by their centred finite
difference approximation over one grid cell. A third order Runge-Kutta
method is used for the time step. In Figure \ref{FigGyroA} the
$v_x$--$v_y$ distribution function is shown for the case of $N_v=30$
and $\Delta t=2\pi\Omega_c/200$ using the flux conservative scheme
with the back substitution algorithm for integration of the
characteristics. The $z$--scaling is arbitrary but identical in the
three sub--plots of Figure \ref{FigGyroA}. Initially (a), the
distribution function is set to the shifted Maxwellian. At $t=2\pi
\Omega_c$ (b), the peak has performed one full gyration. A slight
dissipation can be observed. The dissipation has increased after five
gyration periods $t=10\pi \Omega_c$ (c). With the above time step, this
correspond to a total of 1000 time steps.

\begin{figure}
\begin{center}
\includegraphics[width=10cm]{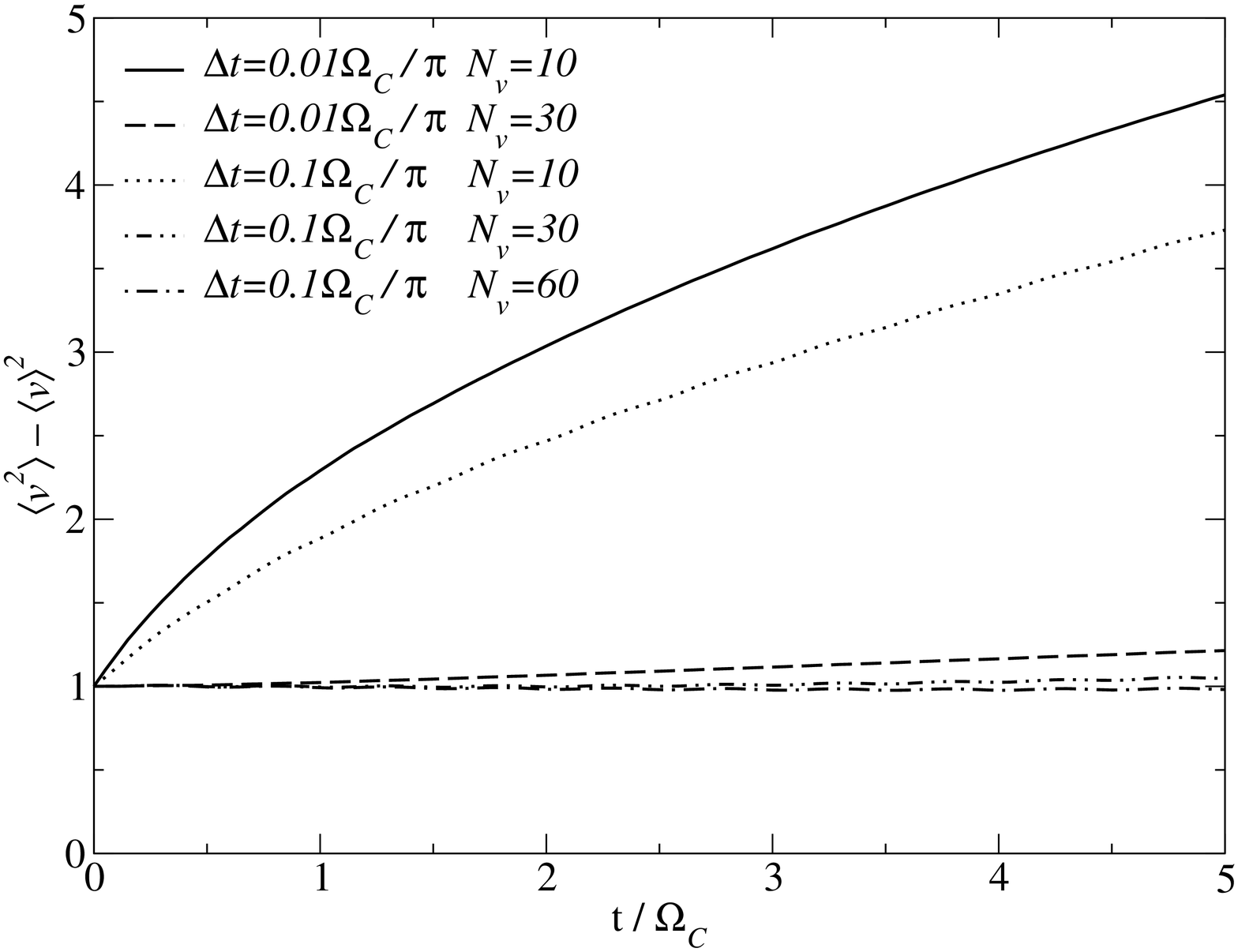}
\includegraphics[width=10cm]{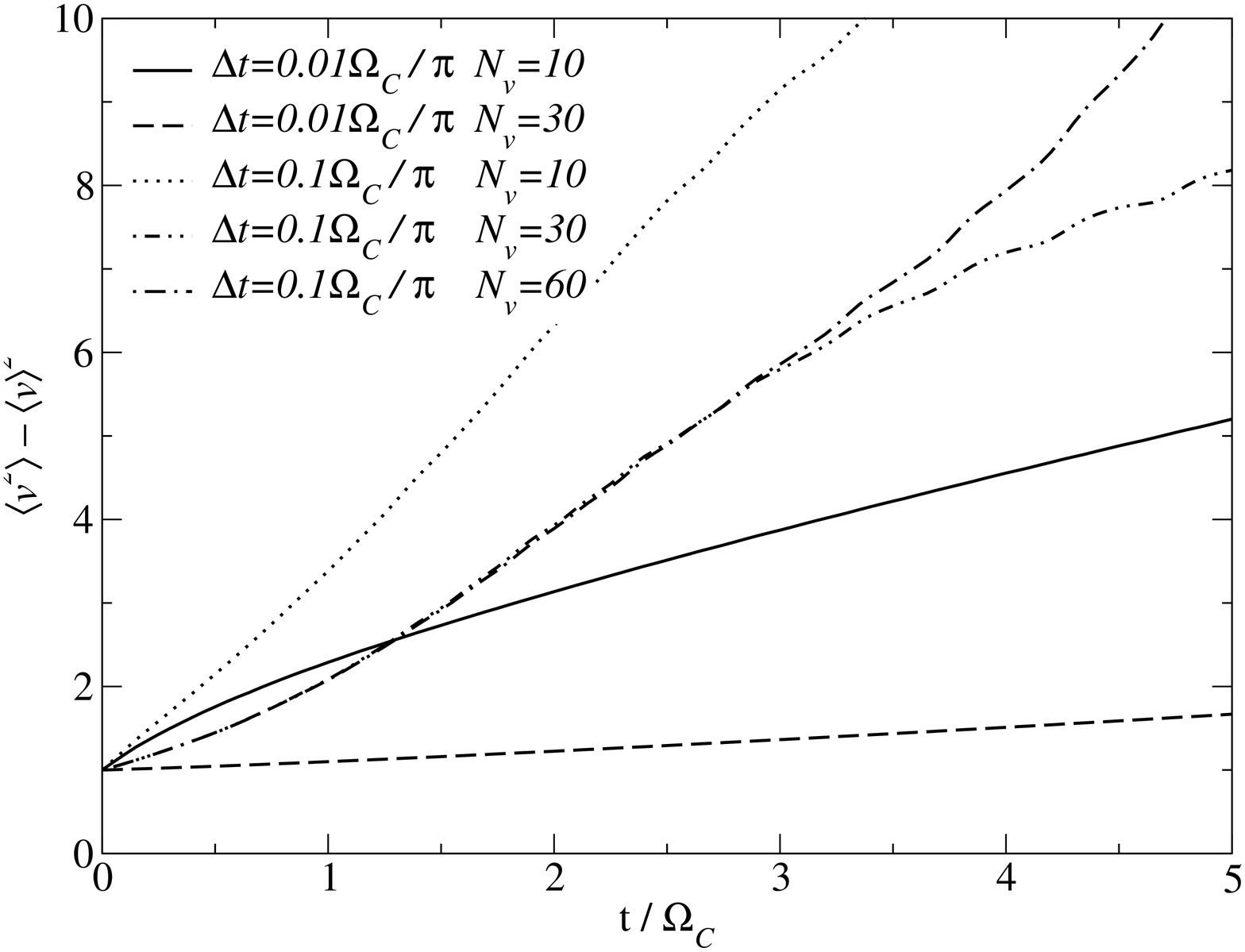}
\end{center}
\caption{The thermal energy of the distribution function $\langle v^2
\rangle -\langle v \rangle^2$ during five gyro--periods for different
time--steps and different phase space resolutions. The upper panel shows the
curves resulting from the back--substitution algorithm, the lower panel
those resulting from the time splitting algorithm. No limiting from above was
applied \label{FigEnergies}}
\end{figure}

The dissipation in phase space can be quantified by taking the second
moment of the distribution function $\langle (v  -\langle v \rangle)^2
\rangle = \langle v^2 \rangle -\langle v \rangle^2$ which represents the
thermal energy in the system. These have been calculated over a time of
five gyro--periods for different simulation time steps, different grid
resolutions of the phase space, and different time splitting algorithms. In
Figure \ref{FigEnergies} the thermal energies are shown for both the
back--substitution algorithm (upper panel) and for the time splitting algorithm
(lower panel), both taken without limiting the distribution function from
above. For a small time step $\Delta t = 0.01 \Omega_c /\pi$, i.e.\ a
time--resolution of 200 time steps per gyration, calculations using
$N_v=10$ and $N_v=30$ grid points per velocity dimension have been carried
out. In the case of $N_v=10$  and using the back--substitution algorithm,
the thermal energy rises more than a factor of 4.5 during the simulated
time period. The amount of diffusion can be reduced considerably by
increasing the grid resolution. At $N_v=30$, the increase in thermal energy
reduces to around 20\% over the simulated five gyro--periods. 

For $N_v=10$, an increase in the time step by a factor of 10 does not
modify the results by a great amount. The already large diffusion values
remain similar although somewhat lower. In contrast, for the $N_v=30$ case,
the thermal energy at the end of the simulation is risen only by less than
5\% from the starting value, for a time step of $\Delta t = 0.1 \Omega_c
/\pi$, i.e.\ 20 time steps per gyration. This means that an increase in
the time steps leads to better results for the thermal energy. An increase
in grid resolution further improves the result and reduces the error to
less than 2\%.

The results using the time splitting scheme (lower panel of Figure
\ref{FigEnergies}) differ considerably from the previous results. Again,
the thermal energy in the small time step case with $N_v=10$ rises
strongly. This now gets substantially less accurate by increasing
the time step to 20 steps per gyration. Here, the thermal energy increases
by more than a factor of 12! The best result is achieved by taking a small
time step and a grid resolution of $N_v=30$. Instead of a slight increase
in thermal energy, a slight decrease can be observed here. Increasing the
time step to 20 steps per gyration degrades the results again. A strong
increase in the thermal energy is observed, which is even {\em
worse} for {\em higher} grid resolutions. 

\begin{figure}
\begin{center}
\includegraphics[width=10cm]{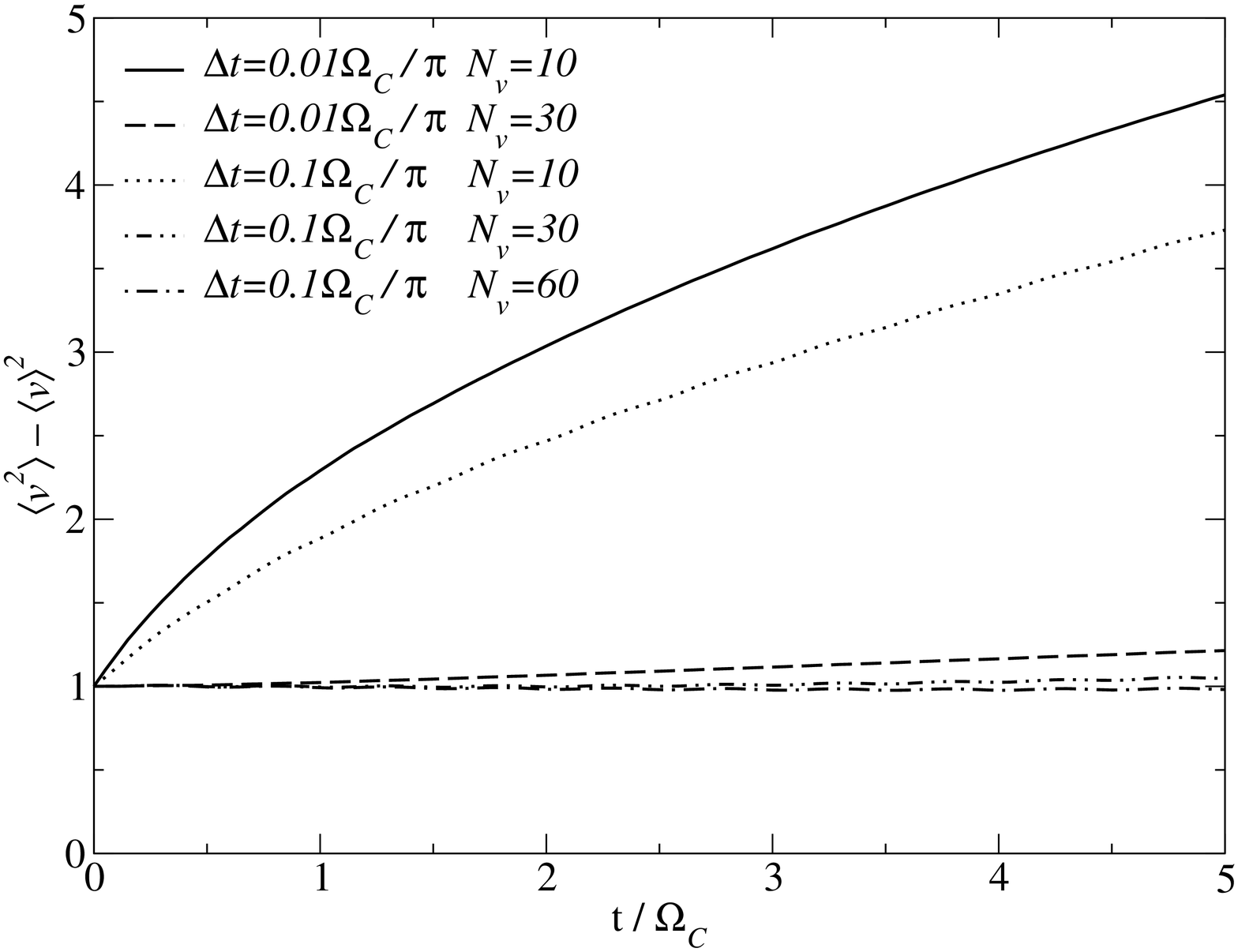}
\includegraphics[width=10cm]{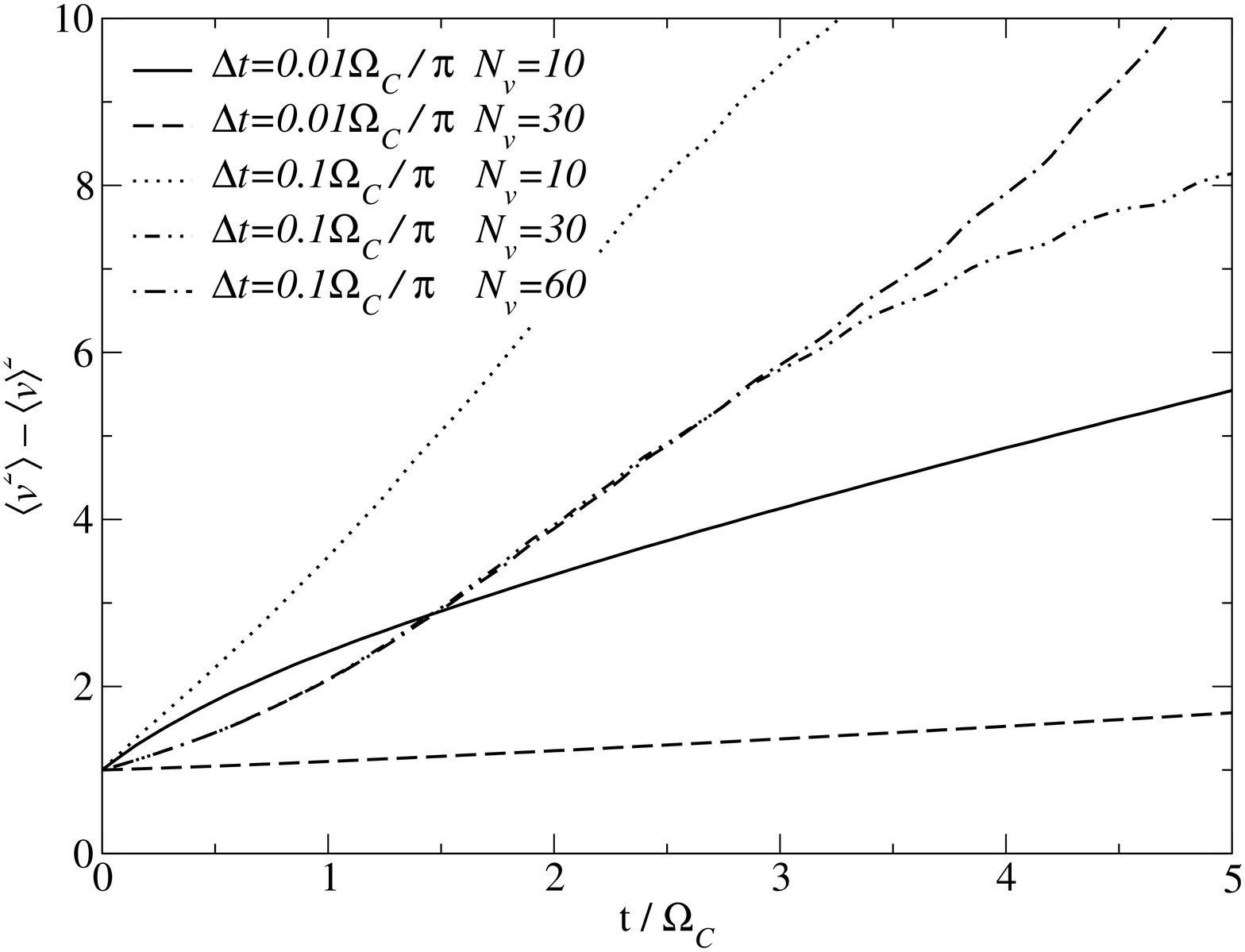}
\end{center}
\caption{The thermal energy of the distribution function $\langle v^2
\rangle -\langle v \rangle^2$ during five gyro--periods for different
time--steps and different phase space resolutions. The upper panel shows the
curves resulting from the back--substitution algorithm, the lower panel
those resulting from the time splitting algorithm. Limiting from above was
applied\label{FigEnergiesLimit}}
\end{figure}

In Figure \ref{FigEnergiesLimit} the same results are shown again, but this
time with the limiting from above switched on. These results differ only
marginally from those obtained without the limiter in place. Note, that the
limiter ensuring positivity of the distribution function is always in
place. Summarising, one finds that the back--substitution method gives
superior results over the time--splitting method. Furthermore in terms of
thermal energy, the back--substitution method rewards larger time steps
(i.e. less computational time) with higher accuracy.

\begin{figure}
\begin{center}
\includegraphics[width=10cm]{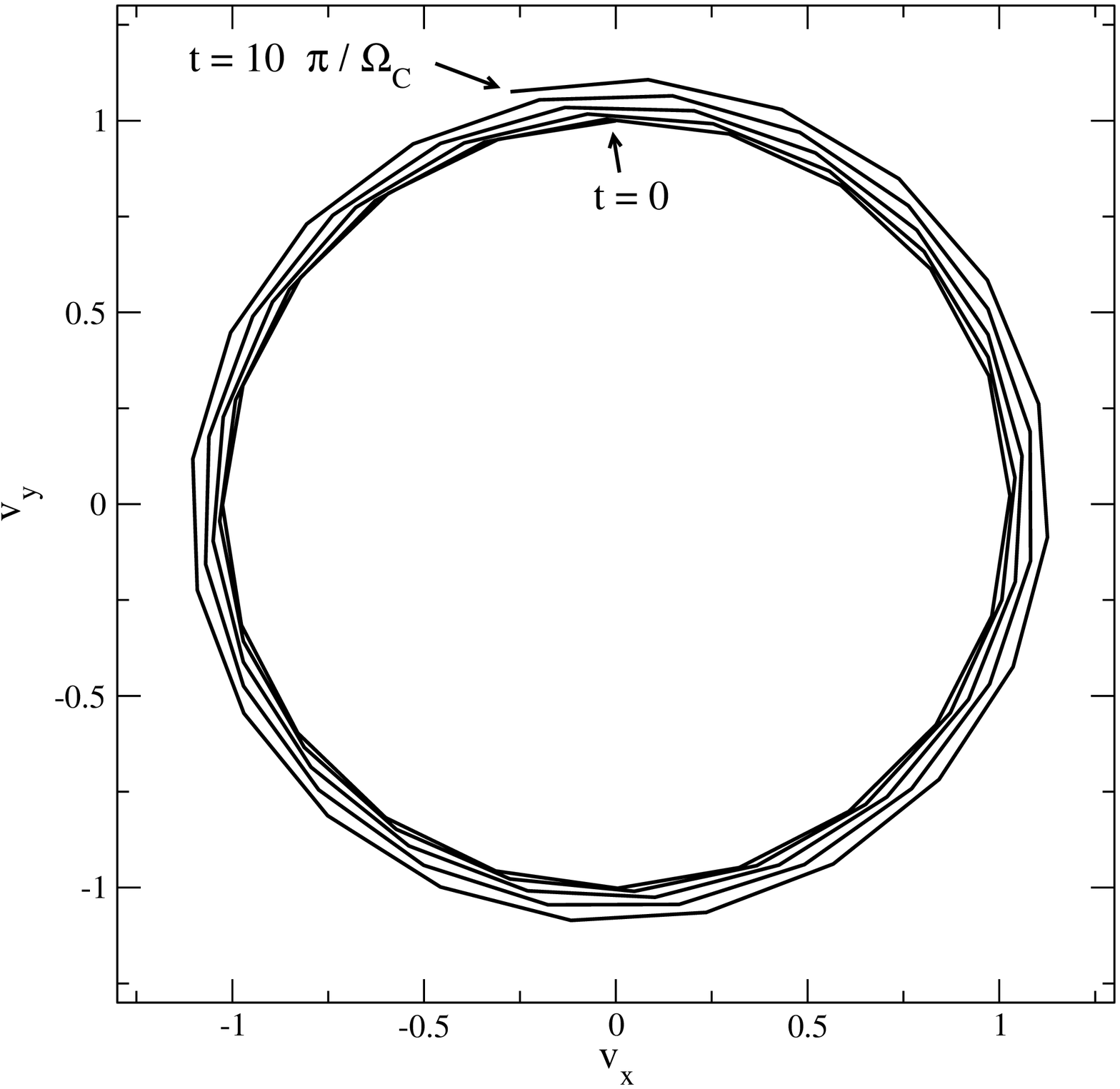}
\end{center}
\caption{The gyro-motion of the Maxwellian peak expressed in $\langle v_y
\rangle$ against $\langle v_x \rangle$ over five gyro--periods. After five
gyro--periods the mean velocity has an error in magnitude and phase from
the starting value.\label{FigGyroMotion}}
\end{figure}

Instead of looking at the thermal energy, which is related to dissipation
in phase space, one can also investigate the gyro--motion of the
Maxwellian peak. From the distribution function the averages $\langle v_x
\rangle$ and $\langle v_y \rangle$ are taken and plotted against each
other through time. In the exact case, this should result in perfectly
circular motion with exactly one turn during a gyration period $2\pi /
\Omega_C$. The numerical Vlasov scheme does not, however, produce exact
results and thus one will observe some form of spiral. An example of this
spiral is plotted in Figure \ref{FigGyroMotion}.  

\begin{center}
\begin{table}[t]
\begin{tabular}{|c|c|c|c|c|c|c|c|}
\hline
        & & \multicolumn{2}{c|}{Back Substitution}  & \multicolumn{2}{c|}{Splitting}&
                        \multicolumn{2}{c|}{Finite Diff.}  \\ \cline{3-6}
$\Delta t$ & $N_v$ &limiter & no limiter & limiter & no limiter & 2nd order & 4th order \\ \hline \hline
$\pi/100$ & $10 $ &  0.8892 & 0.9178 & 1.0843 & 1.0496 & 1.09933  & 1.04747  \\ \hline
$\pi/100$ & $30 $ &  0.9981 & 0.9991 & 1.2029 & 1.2021 & 0.998163 & 0.999867 \\ \hline
$\pi/10$ & $10 $  &  1.1100 & 1.1082 & 2.1824 & 2.0992 & unstable & unstable \\ \hline
$\pi/10$ & $30 $  &  1.1301 & 1.1328 & 2.9553 & 2.9452 & unstable & unstable \\ \hline
$\pi/10$ & $60 $  &  1.1190 & 1.1209 & 4.0757 & 4.0402 & unstable & unstable \\ \hline
\end{tabular}
\caption{Magnitude of the Maxwellian peak after five
gyro--periods for different time--steps, different phase space resolutions
and different integration methods. The magnitudes are normalised to the
initial height of the peak.
\label{TabGyroErrorsMag}}
\end{table}
\end{center}

\begin{table}[t]
\begin{tabular}{|c|c|c|c|c|c|c|c|}
\hline
        & & \multicolumn{2}{c|}{Back Substitution}  & \multicolumn{2}{c|}{Splitting}&
                        \multicolumn{2}{c|}{Finite Diff.}  \\ \cline{3-6}
$\Delta t$ & $N_v$ &limiter & no limiter & limiter & no limiter & 2nd order & 4th order \\ \hline \hline
$\pi/100$ & $10 $ &  0.04214 &  0.06544 &  0.1512  &  0.1363  & 2.05693  & 1.99755  \\ \hline
$\pi/100$ & $30 $ & -0.04178 & -0.03969 & -0.01143 & -0.01341 & 2.07876  & 2.10456  \\ \hline
$\pi/10$ & $10 $  & -0.2500  & -0.2349  &  0.6331  &  0.6002  & unstable & unstable \\ \hline
$\pi/10$ & $30 $  & -0.08965 & -0.09051 &  0.09543 &  0.08768 & unstable & unstable \\ \hline
$\pi/10$ & $60 $  & -0.09232 & -0.09244 &  0.05441 &  0.05011 & unstable & unstable \\ \hline
\end{tabular}
\caption{Phase error of the Maxwellian peak after five
gyro--periods for different time--steps, different phase space resolutions
and different integration methods. 
\label{TabGyroErrorsPhase}}
\end{table}

After some time one can observe errors in both the absolute magnitude of
the velocity and in the phase of the gyro--motion. For the same parameters
as above, the magnitude of the velocity after five gyro--periods is given
in table \ref{TabGyroErrorsMag}, for the various schemes and numerical
parameters. Table \ref{TabGyroErrorsPhase} shows the corresponding phase
errors in radians.

The back--substitution shows better results both for the large and the
small time steps than the time splitting method. The best overall result
is achieved by the back--substitution method in conjunction with a small
time step and a grid resolution of $N_v=30$. Only a marginal phase error
can be observed. There is almost no error in the average velocity. A
decrease of the grid resolution results in a positive phase error and a
decrease of the velocity magnitude error. In contrast, an increase of the
time step causes a rise in the velocity magnitude together with an
increase of the phase error. Here now, an increase in the grid resolution
from 30 to 60 does not improve the results when using a large time step.
For a resolution of 10 and a large time step only the phase error
increases. One can note, that in all these cases the phase error is
negative. This means that the simulated gyro motion always lags behind the
physical gyro motion. 

The errors in phase and velocity magnitude are generally larger when using
the time splitting method. With the large time step the magnitude of the
velocity goes up by approximately 2 to 4 times the initial velocity. The
velocity error is lower when using the small time step; however, for the
high resolution the error is still 20\%. Only the low resolution, large
time step case has a relatively small error in the magnitude. This is
accompanied by a large error in phase. For resolutions 30 and 60
the phase error is better than the one found with the back--substitution
method. The error in the velocity magnitude here is on the other hand,
unacceptable. In both cases one finds that switching on the limiter does
not modify any of these results considerably. 

Finally, comparing this with the results from the finite difference
scheme one finds that the flux conservative approach is far superior
in all cases. Although the finite difference schemes show small
amplitude errors, phase errors are serious. In addition, 
for the large time steps the finite difference scheme
becomes completely unstable (in agreement with the CFL condition).

\subsection{Reconnection\label{SecReconnection}}

A test of the complete Vlasov--Darwin system has been performed on a
magnetic reconnection setup, including full ion and electron dynamics. 
As pointed out in the introduction,
research on magnetic reconnection is still one the most challenging
topics in collisionless plasmas. Kinetic simulations using a Vlasov
code have been carried out, for example by Silin and B\"uchner
\cite{SIL03}.  Since these simulations were carried out using a finite
difference scheme, we will here again present the results of both
finite difference and the flux conservative scheme. We used a
simulation box consisting of 50x100 grid cells in space with a grid
spacing of $\Delta x = 0.1$. The total size of the simulation box is
$L_x=10$ by $L_y=5$. There are 10x10x10 grid cells in velocity
space. Although the resolution in velocity space is --- as shown above
--- insufficient for the gyro motion, especially for the finite
difference scheme, this low resolution is chosen for a number of
reasons. Firstly, it keeps the computational effort to a reasonable
magnitude. Secondly, this corresponds to the phase space resolution
chosen by Wiegelmann and B\"uchner \cite{WIEG01} and thus allows some
comparison. Thirdly, we think that the pure gyro motion test is a very
strict test for the numerical scheme. In the most regions of the
reconnection simulation, the magnetic field is balanced by some
electric field or by the density gradient in such a way that the gyro
motion does not occur the way as presented in the previous section.
The mass ratio between electrons and ions has been chosen to $m_i/m_e
= 16$ and the time step is
\begin{equation}
\Delta t=2.5\cdot 10^{-3} = \frac{1}{25 \Omega_{ce}} ,
\end{equation}
where $\Omega_{ce}$ is the electron Larmor frequency in the unit magnetic
field. Simulations with half this time step have been performed as
convergence test (not shown here) but no substantial deviations have been
found.

\begin{figure}
\begin{center}
\includegraphics[width=12cm]{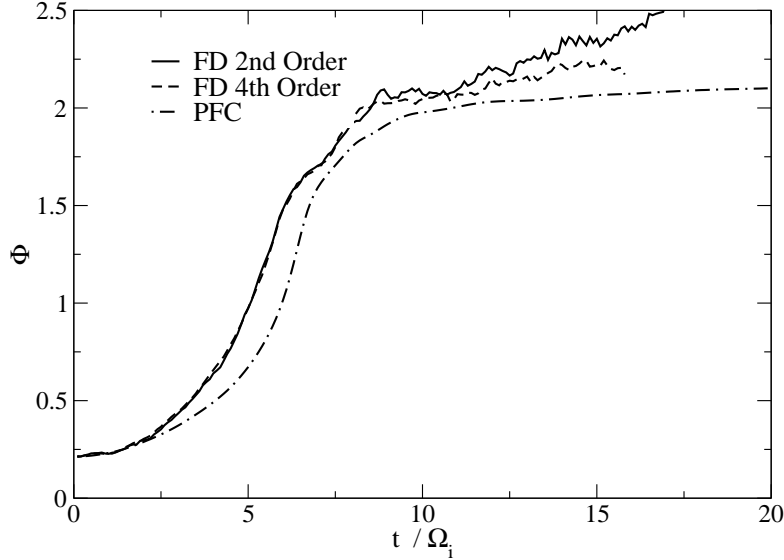}
\end{center}
\caption{The reconnected magnetic flux $\Phi$ over time. The solid line is
the result of the second order finite difference scheme. The dashed line is
the result of a fourth order finite difference scheme. The dashed--dotted
line is the result from the flux conservative scheme.
\label{FigReconFlux}}
\end{figure}

\begin{figure}
\begin{center}
\includegraphics[width=12cm]{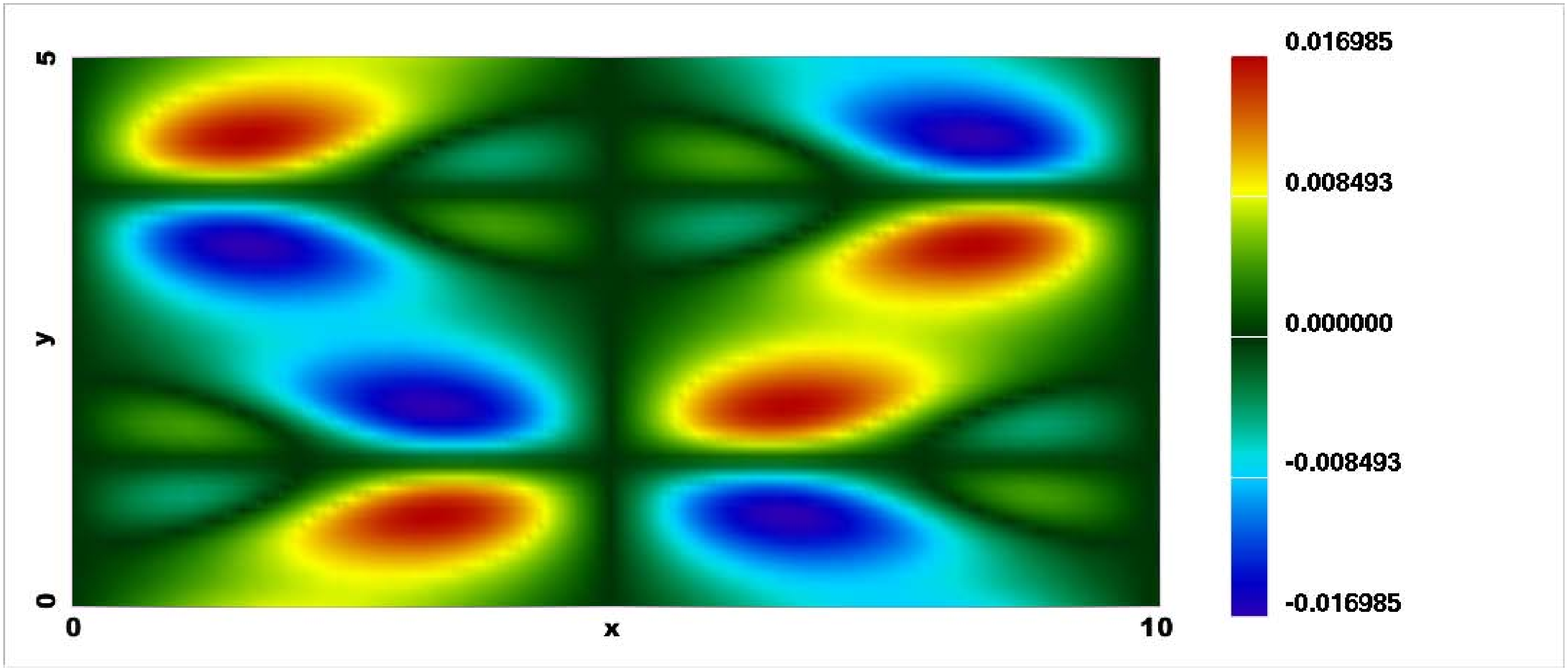}
\includegraphics[width=12cm]{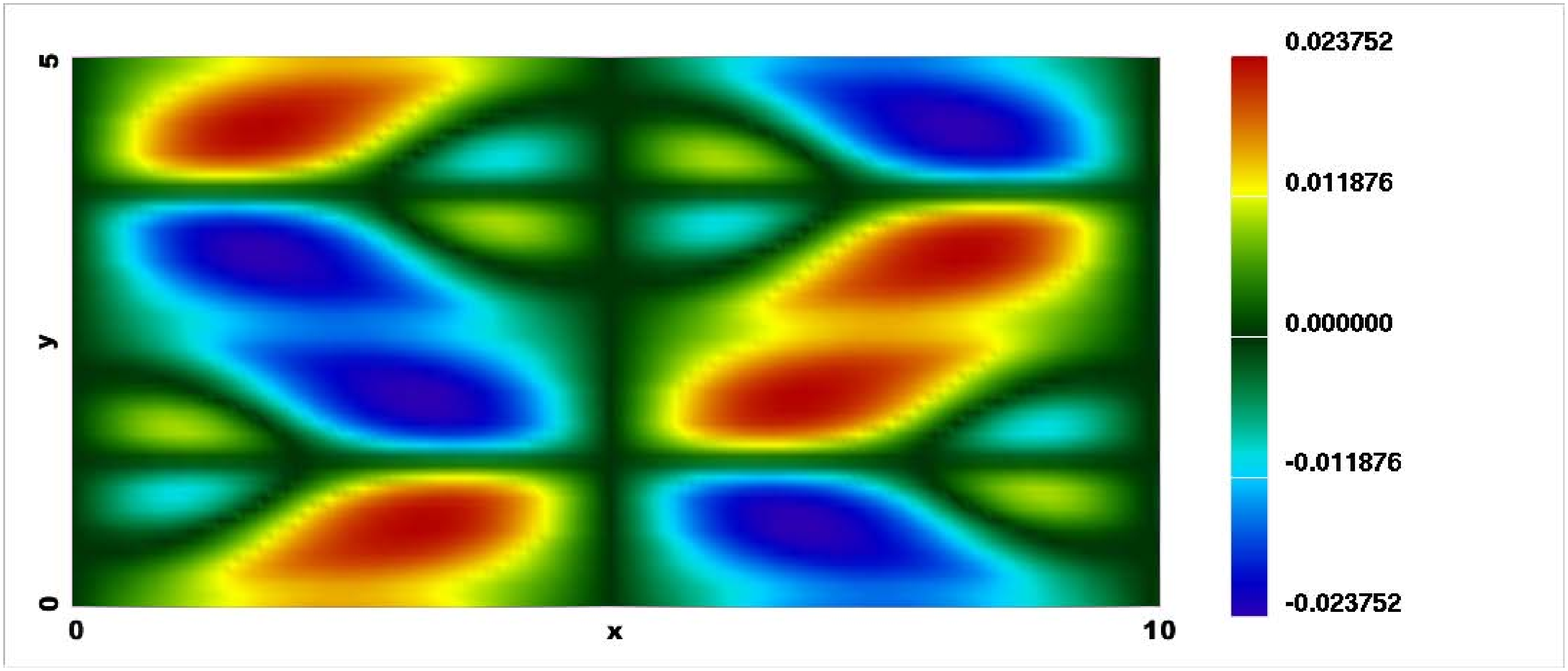}
\includegraphics[width=12cm]{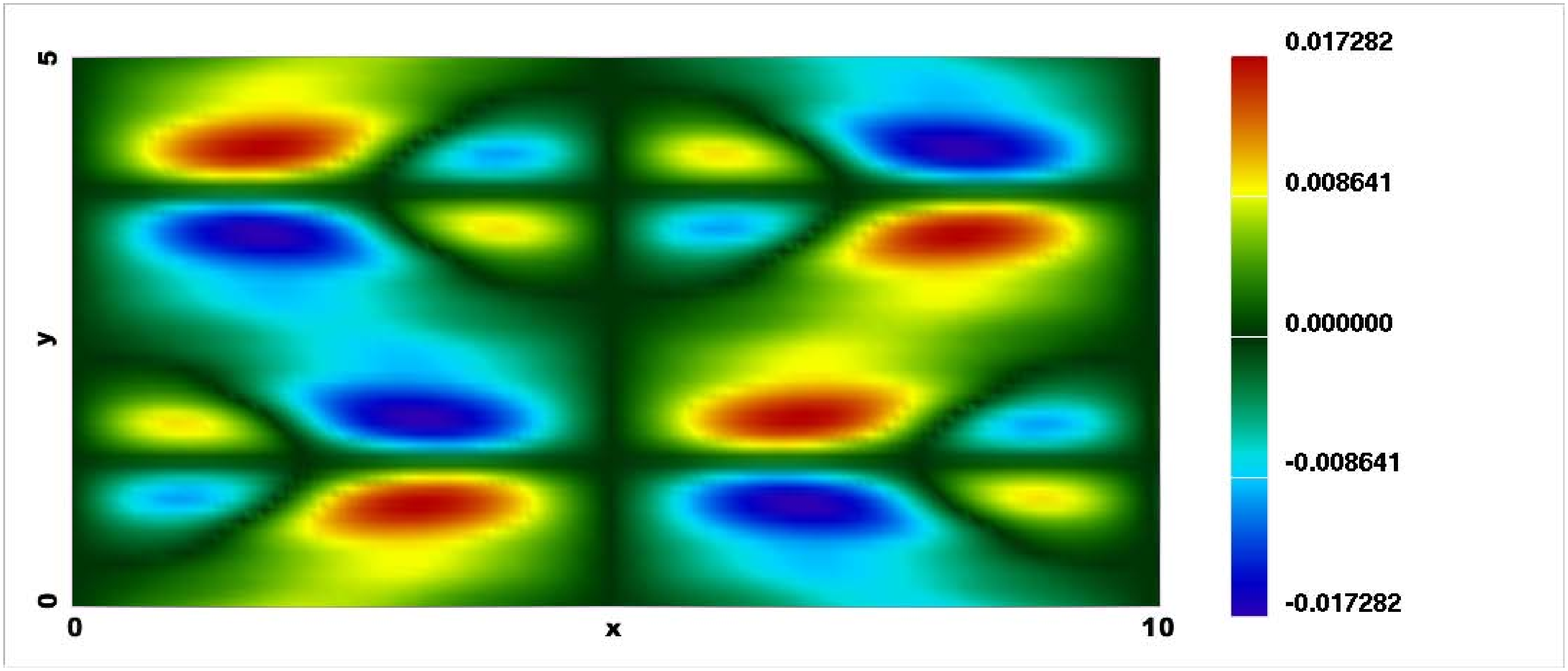}
\end{center}

\caption{The out of plane magnetic field component $B_z$ in the simulation
box at $t=4.1$ using the back--substitution method (upper panel) and at
$t=3.1$ using the finite difference scheme in 2nd order (middle
panel) and 4th order (lower panel).\label{FigReconMag}}

\end{figure}

\begin{figure}
\begin{center}
\includegraphics[width=12cm]{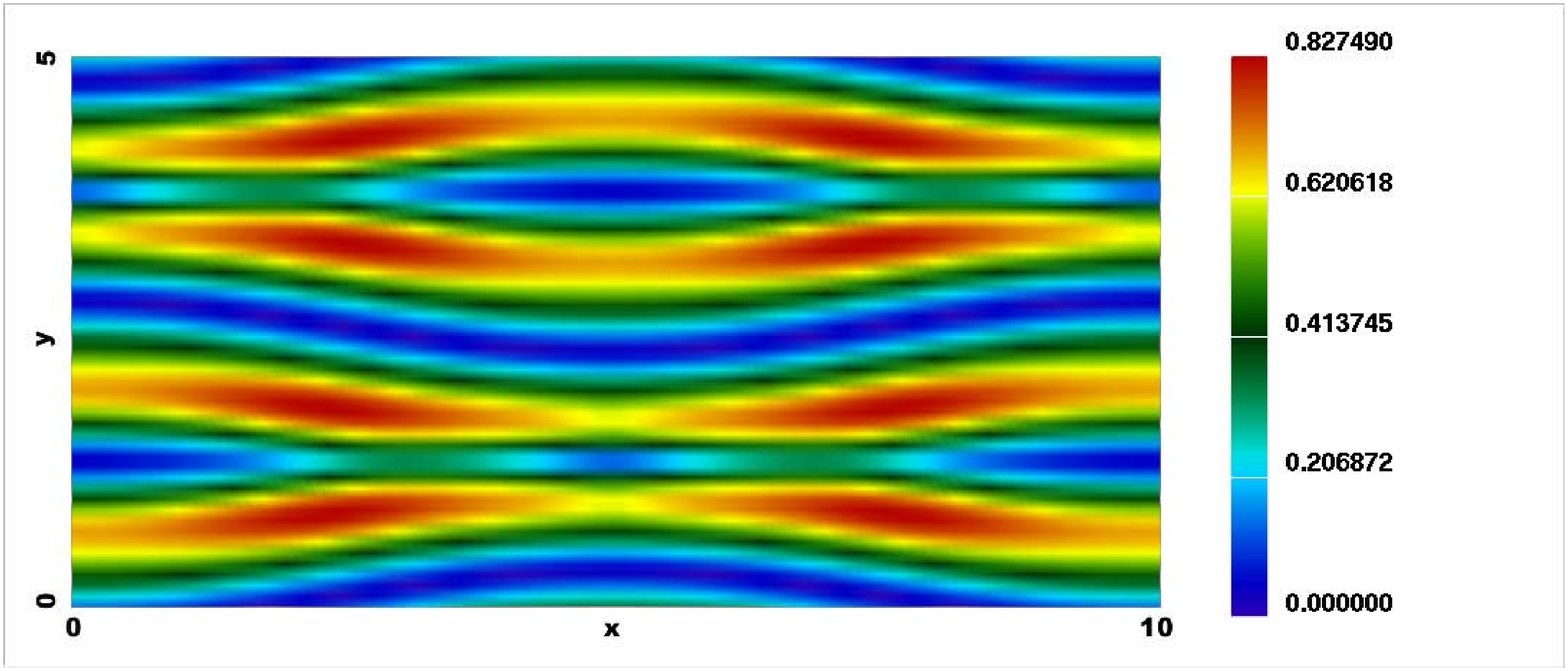}
\includegraphics[width=12cm]{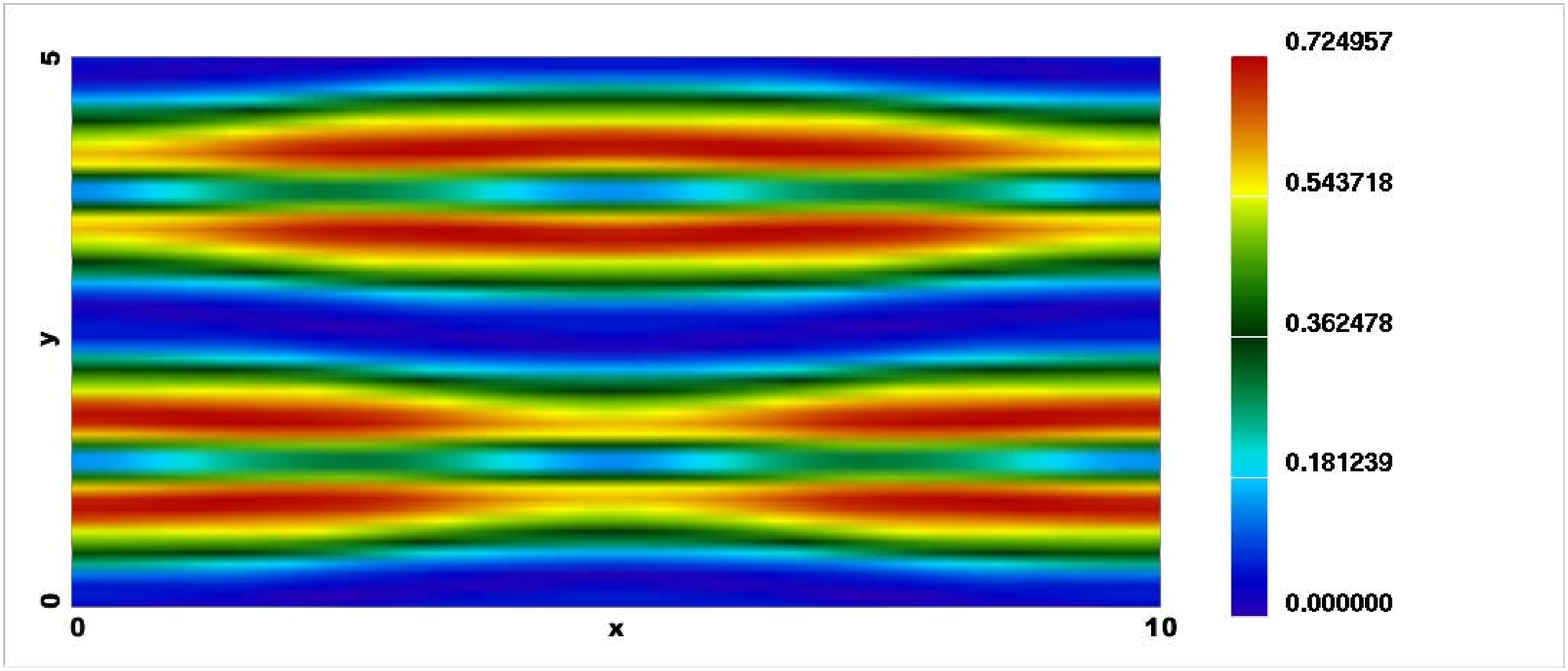}
\includegraphics[width=12cm]{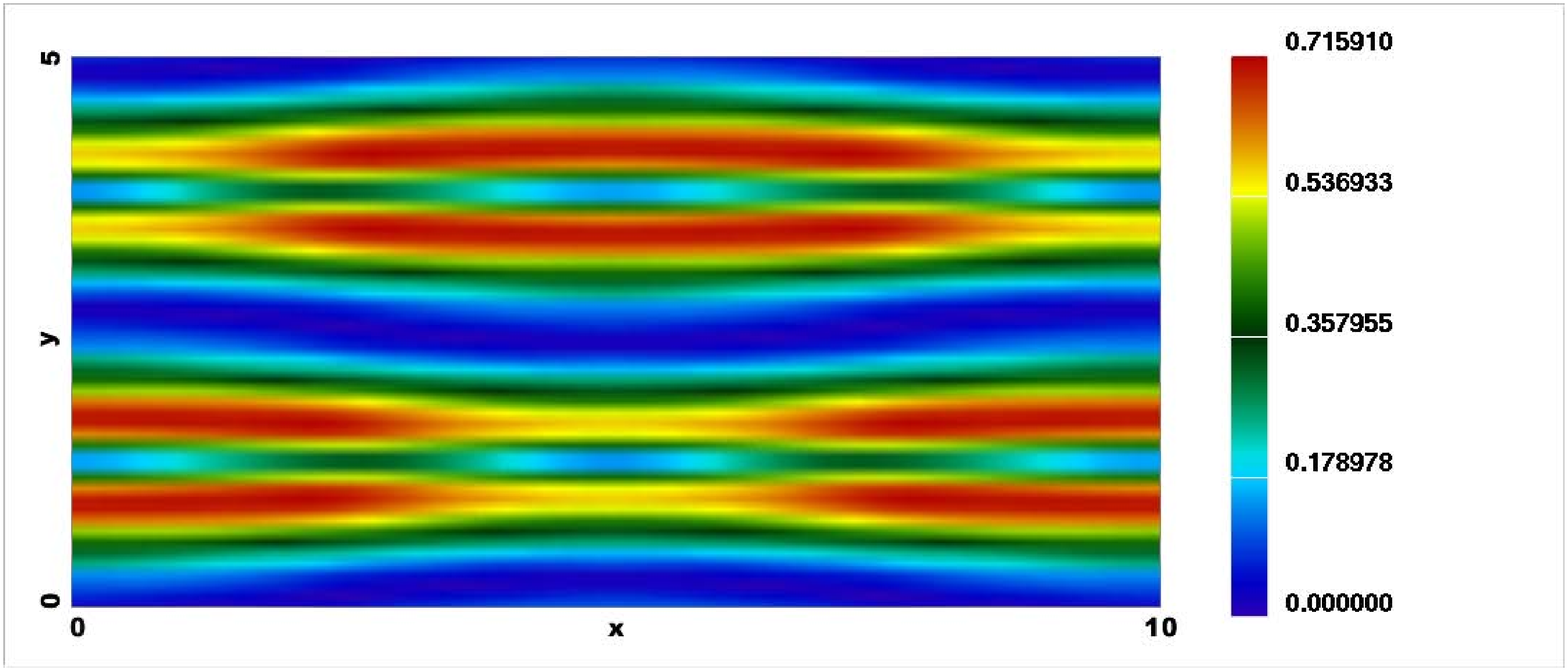}
\end{center}
\caption{The magnitude of the Hall term $|(\mathbf{j}\times\mathbf{B})_z|$ 
in the simulation box at $t=4.1$ using the back--substitution method (upper panel) and at
$t=3.1$ using the finite difference scheme in 2nd order (middle
panel) and 4th order (lower panel).\label{FigReconHall}}
\end{figure}

The initial conditions are chosen as two opposite current sheets. Each
current layer has the well known Harris sheet profile \cite{HARS62}
together with a small perturbation. The distribution function is given by
a shifted Maxwellian
\begin{equation}
f_{i,e}(x,y) = \rho(x,y)\exp\left( \frac{m_j}{2kT_j}
        \left( v_x^2 + v_y^2 + (v_z - v_0^{\pm})^2\right) 
\right) ,
\end{equation}
with $v_0^{\pm} = \pm 1$. The particle density $\rho(x,y)$ is given by
\begin{equation}
\rho(x,y) = \frac{1}{\cosh\left(\left( y-y_0^{\pm}  \right)/\lambda \right)}
\left( 1 + \varepsilon \cos(\frac{2\pi (x - x_0^{\pm})}{L_x}) 
\right) .
\end{equation}
The two Harris sheets, positioned at $y^+ = L_y/4$ and $y^- = 3 L_y/4$,
carry opposite current, and the perturbations have a relative shift along
the $x$--axis of $\pi/L_x$, given by $x^+=0$ and $x^-=1/2$. The
perturbation has a value of $\varepsilon=0.05$. Periodic boundary
conditions are used both in the $x$ and $y$ direction. The X--points will
then develop at $X^+ = (5, 1.25)$ and $X^-=(0, 3.75)$.

Figure \ref{FigReconFlux} shows the temporal evolution of the
reconnected flux. The curves are quite similar keeping in mind that
the resolution is marginal. The major difference is the slightly
earlier onset of reconnection for the finite difference runs.

If one wants to compare the schemes, one has to choose the time for the
different simulations separately such that the reconnected flux is the
same for all simulations. Therefore, we choose a time at the beginning
of the reconnection phase which corresponds to $t=4.1$ for the flux
conservative method and to $t=3.1$ for the finite difference schemes.

Figure \ref{FigReconMag} shows the perpendicular magnetic field $B_z$.
The upper panel shows the result of the flux conservative simulation
while the mid and lower panel display the results of the finite
difference schemes (mid: second order, lower: fourth order). One can
clearly see the quadrupolar structure of the magnetic field component
at the X--point which has been observed in previous simulations of the
reconnection process. There is a slight difference in the distribution
of the magnetic field component between the simulations. One
significant difference is the magnitude of $B_z$. The values are
considerably larger in the second order finite difference scheme than
in the flux conservative scheme and fourth order finite difference
scheme.

In Figure \ref{FigReconHall} the magnitude of the $z$--component of the
Hall-term $|(\mathbf{j} \times \mathbf{B})_z|$ is shown for the
different schemes. This quantity can be taken as a measure for the
importance of the Hall term when comparing to MHD models. All
simulations show qualitatively a similar behaviour.  From the results
of the flux conservative simulation one can see that the
$z$--component of the Hall term is largest outside of the X--points.1

\section{Conclusions\label{SecConclusion}}

A 5 dimensional Vlasov code using the Darwin approximation of Maxwell's
equations has been presented. While in the past a lot of development has
gone into developing numerical schemes for one dimensional electrostatic
Vlasov codes, there is a growing need for kinetic codes that include the
effects of the magnetic field. This is needed for both the simulation of
astrophysical as well as laboratory plasmas. For a large number of
problems the electromagnetic vacuum modes contained in the full set of
Maxwell equations is not of major importance for the physical processes.
These, however, pose severe restrictions on the simulation parameters, due to
the CFL criterion. The Darwin approximation resolves this problem by
eliminating the vacuum modes without reverting to the electrostatic limit.
In contrast to other approximation of Maxwell's equations, the Darwin
approximation can be shown to consistently emerge from an expansion of the
full equations in orders of $v_0^2 / c^2$. The Darwin approximation has
been used extensively in particle in cell simulations but has not found
it's way into Vlasov codes until now. Within the framework of the Darwin
approximation the purely electromagnetic modes are cancelled, but the modes
that rely on the plasma reaction to the magnetic field are retained, such
as the magnetosonic or Alfv\'en modes. 

For the integration of the Vlasov equation in time a recently
developed flux conservative scheme \cite{FIL01} has been used.  The
scheme, which was originally proposed for a one dimensional
electrostatic system had, to be generalised for the higher dimensional
phase space required for the Vlasov--Darwin system. The generalisation
of the one dimensional system is ambiguous and two different schemes
have been compared: the straightforward time splitting scheme
\cite{CAL01} using a generalised time splitting iteration was shown to
suffer from substantial inaccuracies; the back substitution scheme,
proposed in this work, gives far better results not only in terms of an
accurate reproduction of the gyro motion but also concerning the
controllability of the errors. Results were also compared to finite
difference simulations which were shown to be greatly inaccurate in
comparison to the flux conservative scheme.

Finally, a simulation of magnetic reconnection has been performed to test
the full Vlasov--Darwin system. For comparison, the same system has also
been simulated using the finite difference scheme. The finite difference
scheme has been chosen as a reference scheme since this is the most common
approach for simulation magnetic reconnection using a Vlasov code.
Substantial improvements were achieved using the code presented here. The
results especially of derived quantities, did not suffer from numerical
fine scale distortions. It was shown that the results of the scheme
presented here can be treated as much more reliable that those of a finite
difference scheme. 

The code presented here may be used for the investigation of magnetic
reconnection processes and for the simulation of nonrelativistic
collisionless shocks. The main advantages, as compared to PIC simulations,
is the absence of numerical noise and the access to the full distribution
function, especially the high energy tails. In PIC simulations these high
energy tails generally suffer from bad statistics. For this reason it might
be of interest to use Vlasov codes to simulate particle acceleration.

One should note however that, because of the large computational time, the
applicability of Vlasov codes at this time stays restricted to two
dimensional models. Again, in comparison to PIC codes, Vlasov codes are
slower by a factor of a hundred. Despite this comparatively large numerical
effort, Vlasov simulations are still valuable since they complement other
simulation techniques.

\section*{Acknowledgements}

We acknowledge interesting discussions with J. B\"uchner, J. Dreher and
R. Sydora.  This work was supported by the SFB 591 of the Deutsche
Forschungsgesellschaft. Access to the JUMP multiprocessor computer at
Forschungszentrum J\"ulich was made available through project HBO20.

\appendix

\section{Back--Substitution Method for Integration of Characteristics
\label{AppBackSubst}}

Here we present the formulae for the integration of the characteristics in
the velocity space using the back--substitution method. The three
integrations in $v_x$, $v_y$ and $v_z$--direction are carried out
individually. The underlying equations have been presented in section
\ref{SecCharacter} but are in an implicit form for our purpose. Here we
want to give explicit formulae for the three steps
\begin{alignat}{1}
v_x^{n+1} &= v_x^{n+1}(v_x^n, v_y^n, v_z^n) ,\\
v_y^{n+1} &= v_y^{n+1}(v_x^{n+1}, v_y^n, v_z^n) ,\\
v_z^{n+1} &= v_z^{n+1}(v_x^{n+1}, v_y^{n+1}, v_z^n) .
\end{alignat}
Since the bijections between $v^n$ and $v^-$ on one hand and $v^{n+1}$ and
$v^+$ on the other hand are trivial (see eqs (\ref{EqDefVPlus}) and
(\ref{EqDefVMinus})) it is sufficient to formulate the three steps
\begin{alignat}{1}
v_x^+ &= v_x^+(v_x^-, v_y^-, v_z^-) ,\\
v_y^+ &= v_y^+(v_x^+, v_y^-, v_z^-) ,\\
v_z^+ &= v_z^+(v_x^+, v_y^+, v_z^-).
\end{alignat}

By virtue of equations (\ref{EqDefVPrime}) and the $x$--component of
(\ref{EqCalcVPlus}), $v_x$ is already given in the above form. The
integration in $v_x$ is identical to the integration used in the
time splitting algorithm. 

In the second integration step $v_y^+$ has to be determined from $(v_x^+,
v_y^-, v_z^-)$. The $y$--component of equation (\ref{EqCalcVPlus}) gives
$v_y^+ = v_y^+(v_x^-, v_y^-, v_z^-)$. However, the $x$--component of that
equation can now be solved for $v_x^-$ to give
\begin{equation}
v_x^- = v_x^-(v_x^+, v_y^-, v_z^-) 
= \frac{1}{A_x}\left(  
        v_x^+ - v_y^- (s_z+s_y t_x) - v_z^- (s_y-s_z t_x)
\right) ,
\end{equation}
with
\begin{equation}
A_x = 1-s_y t_y - s_z t_z
\end{equation}
With this we get
\begin{equation}
v_y^+(v_x^+, v_y^-, v_z^-) = v_y^+(v_x^-(v_x^+, v_y^-, v_z^-) , v_y^-,
v_z^-) .
\end{equation}

In the same manner the $z$--component of equation (\ref{EqCalcVPlus}) gives
$v_z^+ = v_z^+(v_x^-, v_y^-, v_z^-)$. Then the $x$ and the $y$--components
of that equation are used to solve for $v_x^-$ and $v_y^-$

\begin{alignat}{1}
v_x^- &= \frac{(1-s_x t_x - s_z t_z) \left(v_x^+ + (s_y-s_z t_x) v_z^-\right) 
        - (s_z + s_y t_x) \left(v_y^+ - (s_x+s_z t_y) v_z^-\right)}{(1-s_y t_y - s_z t_z) (1-s_x t_x - s_z t_z) 
                                + (s_z + s_y t_x) (s_z - s_x t_y)} ,\\
v_y^- &= \frac{(s_z - s_x t_y) \left(v_x^+ + (s_y-s_z t_x) v_z^-\right) 
        + (1-s_y t_y - s_z t_z) \left(v_y^+ - (s_x+s_z t_y) v_z^-\right)}{(1-s_y t_y - s_z t_z) (1-s_x t_x - s_z t_z) 
                                + (s_z + s_y t_x) (s_z - s_x t_y)} .
\end{alignat}
Inserting this into $v_z^+(v_x^-, v_y^-, v_z^-)$ then provides the
expression for $v_z^+ = v_z^+(v_x^+, v_y^+, v_z^-)$.

%\label{}

%\bibliographystyle{elsart-num}
%\bibliography{literatur}

\end{document}